\documentclass[a4paper]{article}

\usepackage{jheppub} 
                     
\usepackage{graphicx,color}
 \usepackage{bm}
   \usepackage{amsmath}
    \usepackage{amssymb}    
     \usepackage{pifont}
 \usepackage{stmaryrd}

\usepackage{standalone}
\usepackage{slashed}
\usepackage{multirow}
\usepackage{makecell}
\usepackage{tabu}
\usepackage{hhline}
\usepackage{colortbl}

\newcommand{\Ds}{\displaystyle}

\newcommand{\nn}{\nonumber}

\newcommand{\Tr}{\mathrm{Tr}}
\newcommand{\sign}{\text{sign}}

\newcommand{\ot}{\leftarrow}

\renewcommand{\(}{\left(}
\renewcommand{\)}{\right)}
\renewcommand{\[}{\left[}
\renewcommand{\]}{\right]}

\renewcommand{\vec}[1]{\bm{#1}}
\newcommand{\fnot}[1]{\slashed{#1}}

\title{Angular distributions of Drell-Yan leptons in the TMD factorization approach}

\author{Sara Pilo\~neta,}

\author{Alexey Vladimirov}

\affiliation{Departamento de F\'isica Te\'orica \& IPARCOS, Universidad Complutense de Madrid, E-28040 Madrid, Spain}

\emailAdd{sarapilo@ucm.es}
\emailAdd{alexeyvl@ucm.es}

\preprint{IPARCOS-UCM-24-037}

\abstract{
We present a comprehensive study of the angular structure functions for Drell-Yan leptons in $Z/\gamma$-boson production within the framework of the transverse momentum dependent (TMD) factorization theorem, including kinematic power corrections (KPCs). We find good agreement with the data in the applicability region of the TMD factorization theorem. The inclusion of KPCs allows us to describe all angular coefficients in a frame-independent manner using only the leading-twist TMD distributions: the unpolarized and the Boer-Mulders functions. The value of the Boer-Mulders function is determined using the ATLAS measurement of the $A_2$ angular coefficient. The analysis is performed at N$^4$LL perturbative order. Additionally, we discuss the technical implementation and impact of KPCs on the phenomenology of TMD distributions.
}

\begin{document} 
\allowdisplaybreaks
\maketitle 

\section{Introduction}

The physics of the production of transverse momentum gauge bosons provides a wealth of information about many aspects of QCD and the Standard Model. There are nine angular structure functions for unpolarized scattering, each sensitive to different aspects of parton dynamics, which were derived long ago within the collinear factorization formalism \cite{Collins:1977iv, Arnold:1988dp, Chiappetta:1986yg, Gonsalves:1989ar, Mirkes:1992hu, Mirkes:1994eb}. Nowadays, it is well-established that this approach yields a good description of these functions at large and moderate values of $q_T$ \cite{Karlberg:2014qua, Lambertsen:2016wgj, Gauld:2017tww, Gauld:2021pkr}. On the contrary, there have been limited analyses focusing on the exploration of angular coefficients in the low-$q_T$ region \cite{Barone:2010gk, Lu:2011mz}. This work is the first systematic (albeit in some aspects still preliminary) study of angular structure functions in $Z/\gamma$-boson production using the transverse momentum dependent (TMD) factorization theorem framework.

Within the QCD factorization method, the various angular structure functions of the Drell-Yan (DY) reaction have distinct origins and leading non-vanishing orders. Only four of them are present at the leading power (LP) in the TMD factorization framework \cite{Tangerman:1994eh, Arnold:2008kf}, while the rest are generated by power corrections. TMD factorization has an intricate structure of power corrections \cite{Vladimirov:2023aot} due to its inherent multi-dimensionality. Different types of power corrections are dominant in different kinematic regimes. In particular, in the natural TMD regime (where $q_T \sim \Lambda_{\text{QCD}}\ll Q$), the kinematic power corrections (KPCs) are the largest, while other types of corrections are much smaller. Consequently, TMD factorization with resummed KPCs, as derived in ref.~\cite{Vladimirov:2023aot}, provides a good approximation for power-suppressed structure functions. KPCs generate non-vanishing contributions to all (except one) angular coefficients in the DY reaction, making them the natural objects to test this novel type of factorization formula.

In many aspects, KPCs are an extension of the LP term. They share the same perturbative and non-perturbative content as the LP term, but fulfill the fundamental properties broken by the LP approximation (such as the charge conservation and the frame invariance). Thus, there are no extra TMD distributions in addition to the familiar unpolarized and Boer-Mulders functions, which represent the TMD structure of the unpolarized hadron at LP. The unpolarized TMD parton distribution function (TMDPDF) has been studied in detail in recent years; see, for instance, refs.~\cite{Bacchetta:2019sam, Bertone:2019nxa, Scimemi:2019cmh, Bacchetta:2022awv, Bury:2022czx, Moos:2023yfa}, and is well known. In this work, we use the so-called ART23 extraction for unpolarized TMDPDF \cite{Moos:2023yfa} (with some minimal modifications described in sec.~\ref{sec:NP-setup}). In contrast, the Boer-Mulders function is practically unknown. It has been extracted in refs.~\cite{Barone:2009hw, Barone:2010gk} based on the analysis of the $\cos2\phi$ distribution at low energy. However, these studies are somewhat limited because they are based on the tree-order approximation and data that do not entirely belong to the TMD factorization region. Therefore, we estimate the Boer-Mulders function using the angular distribution $A_2$ measured at the LHC with the N$^3$LO perturbative input, similar to the state-of-the-art analyses of unpolarized TMD data \cite{Bacchetta:2022awv, Moos:2023yfa}.

This work aims to investigate two main problems: first, the feasibility of using the TMD factorization with KPCs, and second, the possibility of describing the DY angular coefficients within this approach. We reach positive conclusions for both questions, although we also point out tensions between theory and data in some regimes (which, however, are not yet conclusive). Since this project is the first application of the TMD-with-KPCs factorization theorem, we have faced many technical complications to be resolved. The numerical implementation is made on the base of \texttt{artemide} \cite{artemide}, which received a massive update. 

The article is organized as follows. In section \ref{sec:theory}, we review the TMD-with-KPCs factorization formula and derive the theoretical expressions for the angular structure functions, which are discussed in sec.~\ref{sec:theory:structure-functions}. Section \ref{sec:setup} is devoted to the description of the practical setup, such as the orders of the perturbative expressions, the non-perturbative models, and other relevant details. The actual comparison of the data with the theory predictions is presented in section \ref{sec:practice}. Additionally, we also prepare several appendices that highlight particular technical details of our implementation. Appendix \ref{app:couplings} describes the electro-weak input, while appendix \ref{sec:implementation} is dedicated to the practical realization of the cut-factors (appendix \ref{app:cut-factors}), the wide-range Fourier transformation algorithm (appendix \ref{app:Fourier}) and the momentum-space convolution (appendix \ref{app:convolution}).

\section{Theory}
\label{sec:theory}

In this work we consider the production of a neutral gauge boson decaying into a Drell-Yan lepton pair
\begin{eqnarray}\label{reaction}
h_1(p_1)+h_2(p_2)~\to~ \gamma^*/Z(q)+X ~\to~ \ell^-(l)+\ell^+(l') +X,
\end{eqnarray}
where in brackets we indicate the momenta of the respective particles. In the following, we assume that leptons and hadrons are massless, i.e., $l^2=l'^2=p_{1,2}^2=0$. In the leading electro-weak order, the differential cross-section of the above reaction (\ref{reaction}) takes the form
\begin{eqnarray}\label{dsigma:0}
d\sigma=\frac{2\alpha_{\text{em}}^2}{s} \frac{d^3l}{2E}\frac{d^3l'}{2E'}\sum_{GG'}L_{\mu\nu}^{GG'}W_{GG'}^{\mu\nu}\Delta^*_G(q)\Delta_{G'}(q).
\end{eqnarray}
Here, $\alpha_{\text{em}}=e^2/4\pi$ is the QED coupling constant, $s=(p_1+p_2)^2$ is the center-of-mass energy, the index $G$ runs over the gauge boson type, and $\Delta_G$ is the propagator of the corresponding gauge boson.

The kinematics of the process is defined by the gauge-boson momentum $q^\mu$, with $q^2=Q^2$, and the hadrons momenta $p_1^\mu$ and $p_2^\mu$, with $p_{1,2}^2=0$. We adopt the following convention for the variables involved. The hadrons momenta define two collinear directions
\begin{eqnarray}
p_1^\mu=\bar n^\mu p_1^+,\qquad p_2^\mu=n^\mu p_2^-,
\end{eqnarray}
where $n^\mu$ and $\bar n^\mu$ are two light-like vectors normalized as $(n\bar n)=1$. We utilize the usual notation for the light-cone decomposition of a generic vector
\begin{eqnarray}
v^\mu=\bar n^\mu v^++n^\mu v^-+v_T^\mu,
\end{eqnarray}
where $v^+ = (nv)$, $v^- = (\bar nv)$, and $v_T$ is the transverse component orthogonal to the $(n,\bar n)$-plane, $v_T^\mu = g_T^{\mu\nu}v_\nu$, with $g_T^{\mu\nu} = g^{\mu\nu} - n^\mu\bar n^\nu - \bar n^\mu n^\nu$ and $v_T^2<0$. We use the bold-font notation for the transverse vectors which obey the Euclidean scalar product, i.e., $\vec q_T^2=-q_T^2>0$. For future convenience we introduce the variable
\begin{eqnarray}
\tau=\sqrt{2q^+q^-}=\sqrt{Q^2-q_T^2}=\sqrt{Q^2+\vec q_T^2}.
\end{eqnarray}
Lastly, the transverse anti-symmetric tensor is defined as
\begin{eqnarray}
\epsilon_T^{\mu\nu}=\frac{\epsilon^{\mu\nu\alpha\beta}p_{1\alpha} p_{2\beta}}{(p_1p_2)}=\epsilon^{\mu\nu\alpha\beta}\bar n_\alpha n_\beta,
\end{eqnarray}
with $\epsilon^{0123}=+1$.

The hadron and lepton tensors are independent theoretical constructs. The hadron tensor $W^{\mu\nu}$ incorporates all information related to the hadronic structure, and is independent on the type of measurement. On the other hand, the lepton tensor $L^{\mu\nu}$ encapsulates the details of the detected lepton pair, and, consequently, the measurement-related information. There are two main types of measurements to consider: the angular distribution of the lepton pair, and the cross-section integrated in the detector acceptance region (referred to as the fiducial cross-section). These seemly different measurements can be related to each other due to the fact that the lepton tensor is sufficiently simple and the hadron tensor is universal.

In sec.~\ref{sec:W}, we consider the hadron tensor in the framework of the TMD factorization theorem supplemented with KPCs. In sec.~\ref{sec:L} we present a decomposition of the lepton tensor into independent tensors, resulting in the angular distributions and/or fiducial cross-section. Unlike classical considerations, such as those in refs.~\cite{Lam:1978zr, Mirkes:1992hu}, where the hadron tensor is decomposed into basis tensors, we decompose the lepton tensor instead. This choice is motivated by the fact that the lepton matrix element has a simple and exact structure. Meanwhile, the hadron tensor is a complicated object whose properties are driven by the non-perturbative effects rather than by its tensor structure.  

\subsection{Hadron tensor in TMD factorization with kinematic power corrections}
\label{sec:W}

The theory of KPCs for TMD factorization has been developed in ref.~\cite{Vladimirov:2023aot}. In this section we resume the elements of it that are necessary to derive the cross-section of the unpolarized Drell-Yan process. 

The starting point of the derivation is the hadron tensor for the Drell-Yan reaction
\begin{eqnarray}
W_{GG'}^{\mu\nu}=\int \frac{d^4y}{(2\pi)^4}e^{-i(yq)}\sum_X \langle p_1,p_2|J_{G}^{\dagger \mu}(y)|X\rangle \langle X|J_{G'}^\nu(0)|p_1,p_2\rangle.
\end{eqnarray}
Here, $J^\mu_G(x)$ is the electro-weak (EW) current for a boson of type $G$ (which could be $Z$ or $\gamma$ in the present context). Specifically, this current reads
\begin{eqnarray}\label{EW_current}
J^\mu_G(x)=\bar q(x)\gamma_{G}^\mu q(x),\qquad
\gamma_{G}^\mu =g_R^{G}\gamma^\mu(1+\gamma^5)+g_L^{G}\gamma^\mu(1-\gamma^5),
\end{eqnarray}
where $q(x)$ is the quark field and $g^{G}_{R,L}$ are the right and left coupling constants. In this expression, the flavor indices of the quark fields and the couplings, as well as the sum over all flavors, have been omitted for brevity. 

The coupling constants values for $Z$ and $\gamma$-bosons are
\begin{eqnarray}
g_R^{Z}=\frac{-e s_W^2}{2s_Wc_W},\qquad g_L^{Z}=\frac{T_{3}-e s_W^2}{2s_Wc_W},\qquad g^\gamma_{R}=g^\gamma_{L}=\frac{e}{2},
\end{eqnarray}
with $e$ and $T_{3}$ being the charge and the third projection of the iso-spin for the given quark, and $s_W$ and $c_W$ being the sine and cosine of the Weinberg angle. These couplings are related to the vector and axial couplings as
\begin{eqnarray}
v^G=2s_Wc_W(g_R^{G}+g_L^{G}),\qquad a^G=2s_Wc_W(g_L^{G}-g_R^{G}).
\end{eqnarray}

There are several approaches to derive the TMD factorization theorem \cite{Becher:2010tm, Collins:2011zzd, Echevarria:2011epo, Balitsky:2020jzt, Vladimirov:2021hdn}. All of them result into the same LP expression for the hadronic tensor. The derivation of the KPCs is performed using background-field theory techniques and the operator product expansion. In this approach, one starts with the general correlator currents, and explicitly integrates over the high-energy components of the fields using the background-field formulation \cite{Abbott:1980hw, Abbott:1981ke}. The resulting effective operator is systematized with respect to counting rules and evolution properties. Such a direct consideration allows for an easier generalization and interpretation compared to the method-of-regions-type approaches, since one works directly with operators. Details of the application to TMD factorization can be found, for instance, in refs.~\cite{Balitsky:2020jzt, Vladimirov:2021hdn}.

The systematization of the higher power operators is the key part. The operators and their corresponding distributions can be partitioned into independent subsets, which do not mix under either transformations or QCD evolution. In other words, they represent independent non-perturbative functions. Due to this property, the factorized cross-section splits into blocks that are entirely independent at all powers and perturbative orders. The power corrections that follow the LP term in the block are called KPCs to this term (see, e.g., discussions in refs. \cite{Blumlein:2006ia, Belitsky:2010jw, Braun:2011dg}). The KPCs inherit many properties of the leading term in the series, and are responsible for restoring the symmetry properties, such as charge conservation, translation and rotation invariances, etc.

In the case of TMD operators, such systematization is done with the help of the TMD-twist, introduced in ref.~\cite{Vladimirov:2021hdn} and elaborated in refs.~\cite{Rodini:2022wki, Rodini:2023plb}. The LP of the factorization theorem contains only the TMD distributions of twist-two (or formally, TMD-twist-(1,1) \cite{Rodini:2022wki}). Higher power terms incorporate higher-twist distributions and also derivatives of twist-two distributions. The KPCs to the LP term are those terms that contain \textit{only derivatives of twist-two TMD distributions}\footnote{There are also terms that contain only twist-two distributions but accompanied by singular at $b\to 0$ coefficients. These terms form the $q_T/Q$-correction, and are part of the so-called $Y$-term \cite{Collins:2011zzd}. The $q_T/Q$ corrections should be distinguished from the KPCs, since they obey different properties.}. These terms were derived at all powers in ref.~\cite{Vladimirov:2023aot}. The resulting general formula is rather cumbersome (see equation (3.23) in ref.~\cite{Vladimirov:2023aot}), and contains a sum over terms with growing power of ``long'' derivatives, $\partial_\mu-[\partial_\mu \mathcal{D}]\ln\sqrt{\bar \zeta/\zeta}$, where $\mathcal{D}$ is the Collins-Soper kernel and $\zeta$'s are the scales of rapidity separation. In the $\zeta=\bar \zeta$ case, the series simplifies and can be summed, resulting in a rather simple and physical expression, which is used in this work.

The expression for the hadronic tensor in TMD factorization with resummed KPCs reads
\begin{eqnarray}\nn
W_{GG'}^{\mu\nu}&=&\frac{1}{4N_c}C_0\(\frac{Q^2}{\mu^2}\)\int \frac{d^4y}{(2\pi)^4}e^{-i(yq)}
\sum_{a,b}\sum_f\Big[
\Tr\(\gamma^\mu_{G} \overline{\Gamma}_b\gamma_{G'}^\nu \overline{\Gamma}_a\)
\widetilde{\Psi}_{f/p_1}^{[\Gamma_a]}(y;\mu,Q^2)\overline{\widetilde{\Psi}}_{f/p_2}^{[\Gamma_b]}(y;\mu,Q^2)
\\\label{W_inPOS} &&
\qquad\qquad
+
\Tr\(\gamma^\mu_{G} \overline{\Gamma}_a\gamma_{G'}^\nu \overline{\Gamma}_b\)\overline{\widetilde{\Psi}}_{f/p_1}^{[\Gamma_a]}(y;\mu,Q^2)\widetilde{\Psi}_{f/p_2}^{[\Gamma_b]}(y;\mu,Q^2)\Big]+...~,
\end{eqnarray}
where the dots indicate the contributions with higher-twist distributions and $q_T/Q$ corrections. The functions $\widetilde{\Psi}$ and $\overline{\widetilde{\Psi}}$ represent the TMD-twist-2 part of the correlators of the quark fields renormalized at the ultraviolet $\mu$ and rapidity $\zeta$ scales, i.e.,
\begin{eqnarray}\label{Psi(y)}
\widetilde{\Psi}^{[\Gamma]}_{f/p_1}(y;\mu,\zeta)&=&\langle p_1|\[\mathcal{Z}(\mu,\zeta)\bar q(y)\frac{\Gamma}{2}q(0)\]_{\text{TMD-tw2}}|p_1\rangle,
\\
\overline{\widetilde{\Psi}}^{[\Gamma]}_{f/p_1}(y;\mu,\zeta)&=&\langle p_1|\[\mathcal{Z}(\mu,\zeta)\Tr[\frac{\Gamma}{2}q(y) \bar q(0)]\]_{\text{TMD-tw2}}|p_1\rangle=-\widetilde{\Psi}^{[\Gamma]}_{f/p_1}(-y;\mu,\zeta).
\end{eqnarray}
Here, $\mathcal{Z}$ represents the renormalization and soft factors composition \cite{Collins:2011zzd,Aybat:2011zv,Echevarria:2012js,Vladimirov:2017ksc}, and $\Gamma$'s are the elements of the Dirac basis used to decompose any matrix $A$ in the spinor representation as
\begin{eqnarray}
A=\frac{1}{2}\sum_{a}\overline{\Gamma}_a A^{[\Gamma_a]},\qquad A^{[\Gamma]}=\frac{1}{2}\Tr\(A \Gamma\).
\end{eqnarray}
The summation $\sum_{a,b}$ runs over all 16 elements of the Dirac basis, the subscript $f$ indicates the flavor of the quark and the coefficient function $C_0$ is the LP coefficient function.

The correlators $\widetilde{\Psi}$ are expressed in terms of ordinary twist-two TMD distributions. The relations are much simpler in the momentum space, and should be derived independently for each element of the Dirac basis. For the hadron moving in the direction $p^\mu=\bar n^\mu p^+$ the correlators are \cite{Vladimirov:2023aot}
\begin{eqnarray}\label{Psi->Phi_1}
\widetilde{\Psi}^{[1]}_{f/p}(y;\mu,\zeta)&=&\widetilde{\Psi}^{[\gamma^5]}_{f/p}(y;\mu,\zeta)=0,
\\
\widetilde{\Psi}^{[\gamma^\mu]}_{f/p}(y;\mu,\zeta)&=&2p^+\int d\xi \,d^4 k\,\delta(k^+-\xi p^+)\delta(k^2)e^{i(ky)}k^\mu \Phi_{f/p}^{[\gamma^+]}(\xi,k_T;\mu,\zeta),
\\
\widetilde{\Psi}^{[\gamma^\mu\gamma^5]}_{f/p}(y;\mu,\zeta)&=&2p^+\int d\xi \,d^4 k\,\delta(k^+-\xi p^+)\delta(k^2)e^{i(ky)}k^\mu \Phi_{f/p}^{[\gamma^+\gamma^5]}(\xi,k_T;\mu,\zeta),
\\\label{Psi->Phi_sigma}
\widetilde{\Psi}^{[i\sigma^{\mu\nu}\gamma^5]}_{f/p}(y;\mu,\zeta)&=&2p^+\int d\xi \,d^4 k\,\delta(k^+-\xi p^+)\delta(k^2)e^{i(ky)}\Big(
\\\nn &&
g^{\mu\alpha}_T k^\nu-g^{\nu\alpha}_T k^\mu+\frac{k^\mu n^\nu-n^\mu k^\nu}{k^+}k_T^\alpha\Big)
\Phi^{[i\sigma^{\alpha +}\gamma^5]}_{f/p}(\xi,k_T;\mu,\zeta),
\end{eqnarray}
where $\Phi$ are the ordinary TMD distributions \cite{Mulders:1995dh, Bacchetta:2006tn}, and the index $\alpha$ is transverse. To obtain the expressions for the hadron moving along $p^\mu=n^\mu p^-$, one should replace $n\leftrightarrow\bar n$ in these equations. Note that the equations (\ref{Psi->Phi_1} -- \ref{Psi->Phi_sigma}) involve a five-dimensional integration, while a TMD-distribution depends only on two variables. This representation is constructed intentionally, since this form allows to display the tensor structure in a compact manner, simplifying the subsequent computation.

Substituting (\ref{Psi->Phi_1} -- \ref{Psi->Phi_sigma}) and evaluating traces one gets the complete expression for the hadron tensor. In the present work, we are interested only in the unpolarized contribution. In this particular case, there are only two distributions that contribute: the unpolarized distribution
\begin{eqnarray}
\Phi_{f/p}^{[\gamma^+]}(\xi,k_T;\mu,\zeta)&=&f_{1;f/p}(\xi,k_T;\mu,\zeta)+...,
\end{eqnarray}
and the Boer-Mulders distribution
\begin{eqnarray}
\Phi_{f/p}^{[i\sigma^{\alpha +}\gamma^5]}(\xi,k_T;\mu,\zeta)&=&-\frac{\epsilon^{\alpha \beta}_Tk_\beta}{M}h^\perp_{1;f/p}(\xi,k_T;\mu,\zeta)+...~,
\end{eqnarray}
where $M$ is a constant with dimensions of mass, which we fix as $M=1$GeV. In these formulas the dots indicate the terms proportional to the spin-vector. Consequently, the hadron tensor for unpolarized scattering reads
\begin{eqnarray}\label{W_Tr}
W_{GG'}^{\mu\nu}&=&\frac{p_1^+p_2^-}{N_c}C_0\(\frac{Q^2}{\mu^2}\)
\int d^4k_1 d^4k_2 \delta(k_1^2) \delta(k_2^2)\delta^{(4)}(k_1+k_2-q)
\\\nn &&
\int d\xi_1 d\xi_2\delta(k_1^+-\xi_1p_1^+)\delta(k_2^--\xi_2p_2^-)\sum_q\Big[
\\\nn &&
\Tr\(\gamma^\mu_{G} \fnot k_2\gamma_{G'}^\nu \fnot k_1\)
f_{1,q/h_1}f_{1,\bar q/h_1}
+
\Tr\(\gamma^\mu_{G} \fnot k_1\gamma_{G'}^\nu \fnot k_2\)
f_{1,\bar q/h_1}f_{1,q/h_1}
-\frac{\epsilon_T^{\mu_1\alpha}k_1^{\alpha}k_1^{\nu_1}\epsilon_T^{\mu_2\beta}k_2^{\beta}k_2^{\nu_2}}{M^{2}}
\Big(
\\\nn &&
\Tr\(\gamma^\mu_{G} \sigma^{\mu_2\nu_2}\gamma_{G'}^\nu \sigma^{\mu_1\nu_1}\)h^\perp_{1,q/h_1}h^\perp_{1,\bar q/h_1}
+
\Tr\(\gamma^\mu_{G} \sigma^{\mu_1\nu_1}\gamma_{G'}^\nu \sigma^{\mu_2\nu_2}\)h^\perp_{1,\bar q/h_1}h^\perp_{1,q/h_1}\Big)\Big],
\end{eqnarray}
where we have suppressed the arguments for the product of TMD distributions, that are $(\xi_1,k_1;\mu,Q^2)$ and $(\xi_2,k_2;\mu,Q^2)$, for simplicity.

Moreover, the expression (\ref{W_Tr}) is transverse, i.e.,
\begin{eqnarray}
q_\mu W^{\mu\nu}_{GG'}=q_\nu W^{\mu\nu}_{GG'}=0.
\end{eqnarray}
Indeed, multiplying by $q^\mu=k_1^\mu+k_2^\mu$, the expressions under the traces simplify to the factors $k_1^2$ and $k_2^2$, that are zero due to the delta-functions. This is a manifestation of the conservation of the vector and axial-vector currents in massless QCD.

The hadron tensor (\ref{W_Tr}) is also frame- or reparametrization-invariant. This invariance is a consequence of the fact that the counting rules for the fields are defined ambiguously, and could be modified by power corrections \cite{Manohar:2002fd, Marcantonini:2008qn}. In practice, it implies the invariance of the factorization formula under particular modifications of vectors $n$ and $\bar n$, which obviously holds since (\ref{W_Tr}) does not depend on these vectors.

Finally, the expression (\ref{W_Tr}) exactly reproduces the LP hadron tensor. To demonstrate it, one should integrate over the ``bad components'' of momenta $k$ (i.e., over $k_1^-$ and $k_2^+$) and take the limit $k_1^+,k_2^-\gg k_{1T}, k_{2T}$ (see also the discussion at the end of section \ref{sec:theory:structure-functions}). The limit should be taken under the sign of integration that makes the procedure badly defined, since the integration covers the entire transverse space. Nonetheless, this is the approximation made at LP.

The main technical difference between the LP and KPCs-improved expressions for the hadron tensor is the convolution integral. It is a ten-dimensional integral restricted by eight delta-functions. Consequently, it can be reduced to a two-dimensional integral with a non-trivial integration region. In contrast to the LP convolution, it involves all arguments of TMD distributions and the integration is made in the finite domain. The details of this convolution integral, as well as its practical realization, are presented in appendix \ref{app:convolution}.

\subsection{Lepton tensor}
\label{sec:L}

The Born approximation for the unpolarized lepton tensor reads
\begin{eqnarray}\label{L_0}
L^{\mu\nu}_{GG'}&=&4\[z_{+\ell}^{GG'}(l^\mu l'^\nu+l'^\mu l^\nu-g^{\mu\nu}(ll'))
-iz_{-\ell}^{GG'} \epsilon^{\mu\nu\alpha\beta}l_\alpha l_\beta' \].
\end{eqnarray}
Here, $l$ and $l'$ are the momentum of the negatively and positively charged leptons, respectively. The factors $z^{GG'}_{\pm\ell}$ denote the subsequent electro-weak coupling combinations
\begin{eqnarray}\label{def:z}
z^{GG'}_{+\ell}&=&2(g^R_Gg^R_{G'}+g^L_Gg^L_{G'})=\frac{v^Gv^{G'}+a^{G}a^{G'}}{4s_W^2c_W^2},
\\
z^{GG'}_{-\ell}&=&2(g^R_Gg^R_{G'}-g^L_Gg^L_{G'})=-\frac{v^Ga^{G'}+a^{G}v^{G'}}{4s_W^2c_W^2}, 
\end{eqnarray}
where the subscript $\ell$ is used to indicate that the coupling constants are taken with lepton quantum numbers. The explicit expressions for these couplings, as well as their numerical values, can be found in appendix \ref{app:couplings}.

As mentioned earlier, two types of measurements are discussed in this work: first, the angular distributions, which are weighted coefficients in the angular decomposition of the cross-section, and, second, the cross-section integrated in a fiducial region. Both can be related to each other and are discussed in the following subsections.

\subsubsection{Angular decomposition of the lepton tensor}

The six-dimensional phase space (PS) of the lepton pair is conventionally split into a boson momentum and an lepton-pair angular part as $dPS=d^4q\,d\Omega$, where $q=l+l'$ and $\Omega$ is the solid angle of the leptons in the Collins-Soper (CS) frame \cite{Collins:1977iv}. The relation between the CS-frame and other frames has been discussed in many articles (see, for instance, refs.~\cite{Argyres:1982kg, Boer:2006eq, Arnold:2008kf, Ebert:2020dfc}), where alternative parametrizations of the angular coefficients can also be found. 

The azimuthal angles of the positively charged lepton in the CS-frame are $\theta$ and $\phi$. To express the lepton tensor via these angles, we use the decomposition of the leptons momenta in terms of $q^\mu$, $p_1^\mu$ and $p_2^\mu$ derived in ref.~\cite{Arnold:2008kf}. In particular, denoting
\begin{eqnarray}\label{lepton-momenta}
l^\mu=\frac{q^\mu+\Delta^\mu_l}{2},\qquad l'^\mu=\frac{q^\mu-\Delta_l^\mu}{2},
\end{eqnarray}
one finds that
\begin{eqnarray}
\Delta_l^\mu&=&
\bar n^\mu q^+\frac{Q}{\tau}\(\cos\theta-\frac{Q}{|\vec q_T|}\sin\theta\cos\phi\)
+
n^\mu q^-\frac{Q}{\tau}\(-\cos\theta-\frac{Q}{|\vec q_T|}\sin\theta\cos\phi\)
\\\nn &&
+q^\mu \frac{\tau}{|\vec q_T|}\sin\theta \cos \phi-\tilde q^\mu \frac{Q}{|\vec q_T|} \sin\theta \sin \phi,
\end{eqnarray}
where we have introduced the fourth vector
\begin{eqnarray}
\tilde q_T^\mu=\epsilon^{\mu \nu}_Tq_\nu.
\end{eqnarray}
This parametrization is specially useful because it satisfies that $(q\Delta_l) = 0$ and $\Delta_l^2 = -q^2$.

Substituting the leptons momenta (\ref{lepton-momenta}) into the unpolarized lepton tensor (\ref{L_0}), we obtain a convenient decomposition
\begin{eqnarray}\label{L:def}
L^{\mu\nu}_{GG'}&=&(-Q^2)\(
z^{GG'}_{+\ell}\sum_{n=U,0,1,2,5,6}S_n(\theta,\phi)\mathfrak{L}_n^{\mu\nu}
+
z^{GG'}_{-\ell}\sum_{n=3,4,7}S_n(\theta,\phi)\mathfrak{L}_n^{\mu\nu}\),
\end{eqnarray}
where the $S_n(\theta,\phi)$ variables constitute a set of independent angular structures and $\mathfrak{L}_n^{\mu\nu}$ are some tensors built from the vectors $q$, $p_{1,2}$ and $\tilde q$. These angular structures read
\begin{align}\nn
&S_{U}(\theta,\phi)=1+\cos^2\theta,
&&S_0(\theta,\phi)=\frac{1-3\cos^2\theta}{2},
&&S_1(\theta,\phi)=\sin2\theta\cos\phi,
\\\label{S-def}
&S_2(\theta,\phi)=\frac{1}{2}\sin^2\theta\cos2\phi
&&S_3(\theta,\phi)=\sin\theta \cos\phi,
&&S_4(\theta,\phi)=\cos\theta,
\\\nn
&S_5(\theta,\phi)=\sin^2\theta\sin2\phi
&&S_6(\theta,\phi)=\sin2\theta \sin\phi,
&&S_7(\theta,\phi)=\sin\theta\sin\phi.
\end{align}
Note that, upon the integration over the full solid angle, all structures, except $S_{U}$, vanish
\begin{eqnarray}
\int d\Omega\, S_n(\theta,\phi)=
\left\{\begin{array}{cl}
\Ds\frac{16\pi}{3},& n=U,
\\
0,&\text{otherwise}.
\end{array}\right.
\end{eqnarray}
The expressions for the $\mathfrak{L}_n^{\mu\nu}$ tensors are
\begin{eqnarray}\label{L:U}
\mathfrak{L}^{\mu\nu}_{U}&=&g^{\mu\nu}-\frac{q^\mu q^\nu}{Q^2},
\\\label{L:0}
\mathfrak{L}^{\mu\nu}_{0}&=&-q_+q_-\(\frac{n^\mu}{q_+}-\frac{\bar n^\mu}{q_-}\)\(\frac{n^\nu}{q_+}-\frac{\bar n^\nu}{q_-}\),
\\\label{L:1}
\mathfrak{L}^{\mu\nu}_{1}&=&q_+q_-\frac{Q}{|\vec q_T|}
\[
\(\frac{n^\mu}{q_+}-\frac{q^\mu}{Q^2}\)\(\frac{n^\nu}{q_+}-\frac{q^\nu}{Q^2}\)
-
\(\frac{\bar n^\mu}{q_-}-\frac{q^\mu}{Q^2}\)\(\frac{\bar n^\nu}{q_-}-\frac{q^\nu}{Q^2}\)\]
,
\\\label{L:2}
\mathfrak{L}^{\mu\nu}_{2}&=&2\(g^{\mu\nu}-\frac{q^\mu q^\nu}{Q^2}\)+\(2\frac{Q^2}{\vec q_T^2}-1\)\mathfrak{L}^{\mu\nu}_{0}
\\\nn &&+
4\frac{Q^2}{\vec q_T^2}q_+q_-\[
\(\frac{n^\mu}{q_+}-\frac{q^\mu}{Q^2}\)\(\frac{n^\nu}{q_+}-\frac{q^\nu}{Q^2}\)
+
\(\frac{\bar n^\mu}{q_-}-\frac{q^\mu}{Q^2}\)\(\frac{\bar n^\nu}{q_-}-\frac{q^\nu}{Q^2}\)\]
,
\\\label{L:3}
\mathfrak{L}^{\mu\nu}_{3}&=&i\frac{\tau}{|\vec q_T|}\(\frac{\epsilon^{\mu\nu \alpha \beta}q_\alpha n_\beta}{q_+}+\frac{\epsilon^{\mu\nu \alpha \beta}q_\alpha \bar n_\beta}{q_-}\)
\\\label{L:4}
\mathfrak{L}^{\mu\nu}_{4}&=&i\frac{\tau}{Q}\(\frac{\epsilon^{\mu\nu \alpha \beta}q_\alpha n_\beta}{q_+}-\frac{\epsilon^{\mu\nu \alpha \beta}q_\alpha \bar n_\beta}{q_-}\)
,
\\\label{L:5}
\mathfrak{L}^{\mu\nu}_{5}&=&\frac{-\tau Q}{2\vec q^2_T}\[
\(2\frac{q^\mu}{Q^2}-\frac{n^\mu}{q_+}-\frac{\bar n^\mu}{q_-}\)\tilde q^\nu
+
\tilde q^\mu\(2\frac{q^\nu}{Q^2}-\frac{n^\nu}{q_+}-\frac{\bar n^\nu}{q_-}\)\],
\\\label{L:6}
\mathfrak{L}^{\mu\nu}_{6}&=&\frac{\tau}{2|\vec q_T|}\[
\(\frac{n^\mu}{q_+}-\frac{\bar n^\mu}{q_-}\)\tilde q^\nu
+
\tilde q^\mu\(\frac{n^\nu}{q_+}-\frac{\bar n^\nu}{q_-}\)\],
\\\label{L:7}
\mathfrak{L}^{\mu\nu}_{7}&=&\frac{2i}{Q|\vec q_T|}\epsilon^{\mu\nu\alpha\beta}q_\alpha\tilde q_\beta 
\\\nn &=&
-i \frac{Q\tau^2}{|\vec q_T|}
\[
\(\frac{n^\mu}{q_+}-\frac{q^\mu}{q^2}\)\(\frac{\bar n^\nu}{q_-}-\frac{q^\nu}{q^2}\)
-
\(\frac{\bar n^\mu}{q_-}-\frac{q^\mu}{q^2}\)\(\frac{n^\nu}{q_+}-\frac{q^\nu}{q^2}\)
\].
\end{eqnarray}
These tensors are dimensionless and also transverse, i.e., $q_\mu \mathfrak{L}^{\mu\nu}_n=q_\nu \mathfrak{L}^{\mu\nu}_n=0$. In the literature, they are often presented in terms of the $Z$ boson polarization vectors \cite{Mirkes:1992hu, Ebert:2020dfc, Lyubovitskij:2024civ}. However, we have found it more convenient to write them in terms of the hadron degrees of freedom.

\subsubsection{Lepton tensor with fiducial cuts}

Modern experiments frequently report the angle-integrated cross-section measurements without corrections for the fiducial region. In such cases, these effects must be taken into account theoretically. This can be done through Monte-Carlo generation of the leptonic final state, as detailed in ref.~\cite{Catani:2015vma}, or semi-analytically, using the method proposed in ref.~\cite{Scimemi:2017etj}. Here, we elaborate the semi-analytical approach, and present the general decomposition of the lepton tensor integrated with fiducial cuts.

The experimental restrictions are imposed on the momenta of the individual leptons, and take the form of constraints on the rapidity $\eta$ and on the transverse component $|\vec l_T|$. They can be accumulated in a step-function in the following manner
\begin{eqnarray}\label{Theta-cuts}
\Theta(\text{cuts})=\left\{
\begin{array}{ll}
1,& \text{if }\eta_\ell,\eta_{\ell'},|\vec l_T|,|\vec l'_T|\in \text{fiducial region},
\\
0,& \text{otherwise}.
\end{array}\right.
\end{eqnarray}
Due to the fact that these restrictions leave the vector $q$ intact, it is convenient to perform the subsequent change of variables for the angle-integrated cross-section
\begin{eqnarray}\label{L-cut:definition}
\int_{\Omega} \frac{d^3l}{2E}\frac{d^3l'}{2E'}L_{GG'}^{\mu\nu}\Theta(\text{cuts})=d^4q \int \frac{d^3l}{2E}\frac{d^3l'}{2E'}L_{GG'}^{\mu\nu}\Theta(\text{cuts})\delta^{(4)}(q-l-l')
=d^4q \widehat{L}^{\mu\nu}_{GG'}(\text{cuts}),
\end{eqnarray}
where $\widehat{L}^{\mu\nu}_{GG'}$ is the lepton tensor that incorporates the fiducial cuts. Note that, since the expression for the lepton tensor is known analytically, the integral can also be expressed analytically.

Furthermore, we observe that the lepton tensor with fiducial cuts remains transverse to $q^\mu$,
\begin{eqnarray}
q_\mu\widehat{L}^{\mu\nu}_{GG'}(\text{cuts})=q_\nu\widehat{L}^{\mu\nu}_{GG'}(\text{cuts})=0.
\end{eqnarray}
Additionally, it is a Lorentz tensor that depends only on the vectors $q$, $n$ and $\bar n$ by means of the rapidity and transverse components definition in eqn.~(\ref{Theta-cuts}). Therefore, it can be decomposed into a set of nine independent tensors constructed from these vectors. The natural and suitable choice is the set of tensors (\ref{L:U} -- \ref{L:7}), as they have a convenient physical interpretation. Thus, we have
\begin{eqnarray}\label{L-cut:decomposition}
\widehat{L}^{\mu\nu}_{GG'}(\text{cuts})=\frac{-2\pi Q^2}{3}\(
z^{GG'}_{+\ell}\sum_{n=U,0,1,2,5,6}\mathcal{P}_n(\text{cuts})\mathfrak{L}_n^{\mu\nu}
+
z^{GG'}_{-\ell}\sum_{n=3,4,7}\mathcal{P}_n(\text{cuts})\mathfrak{L}_n^{\mu\nu}\),
\end{eqnarray}
where the common factor is selected such that the cut-factors $\mathcal{P}_n$ are dimensionless, and
\begin{eqnarray}
\mathcal{P}_{U}(\text{no-cuts})=\frac{2}{\pi}\int d^4l\,d^4l'\delta^{(4)}(q-l-l')\delta(l^2)\delta(l'^2)=1.
\end{eqnarray}
The remaining cut-factors disappear in the limit of the absence of cuts
\begin{eqnarray}
\mathcal{P}_{n}(\text{no-cuts})=0,\qquad \text{for }n\neq U.
\end{eqnarray}
Certain components of $\widehat{L}^{\mu\nu}_{GG'}$ can be obtained by contracting it with the $\mathfrak{L}_n^{\mu\nu}$ tensors, yielding an analytic expression for the integrand over the lepton momenta. The details of this procedure, along with the expressions for the cut-factors, are given in appendix \ref{app:cut-factors}.

\begin{figure}
\begin{center}
\includegraphics[width=0.45\textwidth]{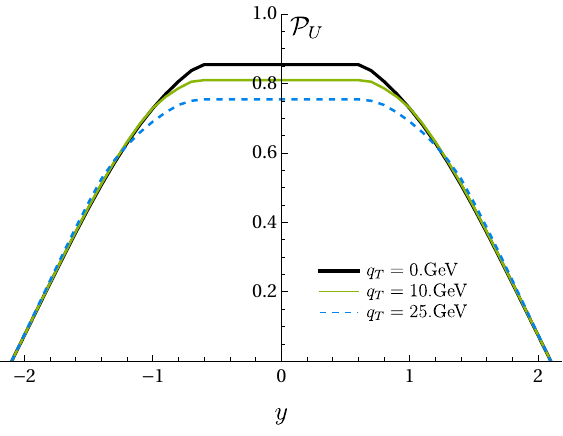}~
\includegraphics[width=0.45\textwidth]{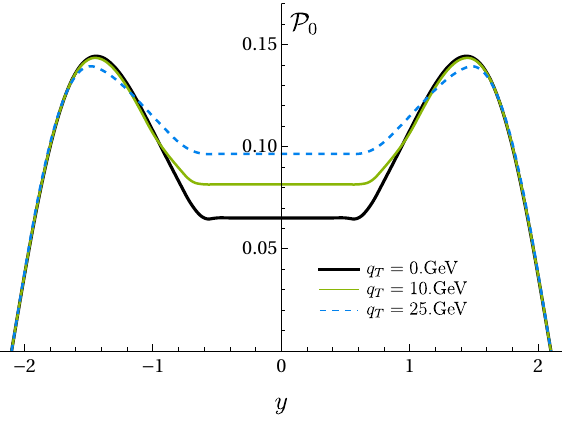}
\end{center}
\caption{\label{fig:cutFactors} The plot for the cut factors $\mathcal{P}_{U,0}$ as a function of $y$ at different values of $q_T$ and $Q=91.$GeV. The cut parameters are $|\eta_{\ell,\ell'}|<2.1$ and $|\vec l_T|,|\vec l'_T|>20$GeV.}
\end{figure}

The majority of the cut-factors are vanishingly small. This happens due to the asymmetry of the integrand with respect to the rapidity or angular variables. In this case, the non-zero values are generated by the asymmetry in the integration domain, which is generally minor. Asymmetric cut conditions (such as those used, e.g., at LHCb \cite{LHCb:2015mad} or CMS \cite{CMS:2016mwa}) can amplify this effect, but it still remains very small. Inspecting various combinations of variables typical for LHC measurements, we have found that the largest contributions naturally come from the factors $\mathcal{P}_U$ and $\mathcal{P}_0$. They are illustrated in fig.\ref{fig:cutFactors} for some typical fiducial region. The factor $\mathcal{P}_1$ (which is 0 at $q_T=0$ and grows almost linearly with $q_T$) can reach values of a few percent. The factor $\mathcal{P}_{2}$ is small; in some regions, it reaches up to $1\%$. The remaining factors are negligible, since $\mathcal{P}_{3,4,5,6,7}<10^{-6}$. 

\begin{figure}
\begin{center}
\includegraphics[width=0.42\textwidth]{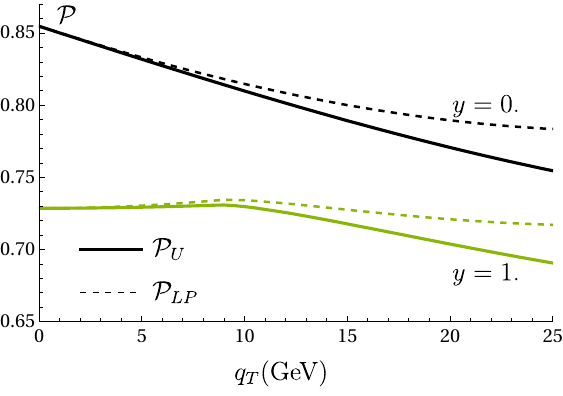}
\end{center}
\caption{\label{fig:PLP_vs_PU} The plot for the cut factors $\mathcal{P}_{U}$ and $\mathcal{P}_{\text{LP}}$ as a function of $q_T$ at different values of $y$ and $Q=91.$GeV. The cut parameters are $|\eta_{\ell,\ell'}|<2.1$ and $|\vec l_T|,|\vec l'_T|>20$GeV.}
\end{figure}

The LP cut factor $\mathcal{P}_{\text{LP}}$ is obtained by convoluting the lepton tensor with $g_T^{\mu\nu}$, which is the tensor part of the LP hadron tensor \cite{Scimemi:2019cmh}. This cut-factor has been used in the ART23 extraction and its predecessors. In fig.\ref{fig:PLP_vs_PU} we compare the values of $\mathcal{P}_{\text{LP}}$ with those of $\mathcal{P}_U$. The difference between them is small at small-$q_T$, but it grows up to $4-5$\% at $q_T\sim 25$GeV. This disparity plays a significant role in the description of the LHC data.

\subsection{Cross-section for the Drell-Yan reaction}
\label{sec:theory:structure-functions}

The standard decomposition of the Drell-Yan cross-section into angular distributions is given by \cite{Mirkes:1994eb}
\begin{eqnarray}\label{def:angular}
\frac{d\sigma}{d^4qd\Omega}&=&\frac{3}{16\pi} \frac{d\sigma}{d^4q}\Big[
(1+\cos^2\theta)+\frac{1-3\cos^2\theta}{2}A_0+\sin2\theta \cos \phi A_1
+\frac{\sin^2\theta \cos2\phi}{2}A_2
\\\nn && 
+\sin\theta \cos\phi A_3+\cos\theta A_4
+\sin^2\theta \sin 2\phi A_5+\sin2\theta \sin\phi A_6+\sin\theta\sin\phi A_7\Big]
\\ \nn &=&\frac{3}{16\pi} \frac{d\sigma}{d^4q}\sum_{n=U,0,...,7}S_n(\theta,\phi)A_n,
\end{eqnarray}
where $A_n$ are the angular distributions, and $A_U=1$. To get their explicit expressions we calculate the differential cross-section (\ref{dsigma:0}) using the hadron (\ref{W_Tr}) and lepton (\ref{L:def}) tensors and utilize that
\begin{eqnarray}
\frac{d^3l}{2E}\frac{d^3l'}{2E'}=\frac{d^4qd\Omega}{8}.
\end{eqnarray}
The computation is purely algebraic and involves evaluating traces and performing various simplifications. This has been accomplished with the help of the \textit{FeynCalc} package \cite{Shtabovenko:2020gxv}. 

The resulting cross-section is conveniently expressed as
\begin{eqnarray}\label{def:sigma_dOmega}
\frac{d\sigma}{d^4qd\Omega}=\frac{3}{16\pi} \sum_{n=U,0,...,7}S_n(\theta,\phi)\Sigma_n,
\end{eqnarray}
such that
\begin{eqnarray}
\Sigma_U=\frac{d\sigma}{d^4q},\qquad A_n=\frac{\Sigma_n}{\Sigma_U}.
\end{eqnarray}
The obtained expressions for the $\Sigma_n$ functions are the following
\begin{eqnarray}\label{def:SU}
\Sigma_U&=&\frac{4\pi \alpha_{\text{em}}^2}{3N_cs}\sum_{q,G,G'}Q^4\Delta_G^*\Delta_{G'}\,z_{+\ell}^{GG'}z_{+q}^{GG'} \,\mathcal{C}[1,f_1f_1],
\\\label{def:S0}
\Sigma_0&=&\frac{4\pi \alpha_{\text{em}}^2}{3N_cs}\sum_{q,G,G'} Q^4\Delta_G^*\Delta_{G'}\Big\{
z_{+\ell}^{GG'}z_{+q}^{GG'}\mathcal{C}[\frac{(\vec k_1-\vec k_2)^2}{Q^2}-\frac{(\vec k_1^2-\vec k^2_2)^2}{\tau^2Q^2},f_1f_1]
\\\nn &&
+z_{+\ell}^{GG'}r_{+q}^{GG'}\mathcal{C}[\frac{(\vec k_1-\vec k_2)^2(\vec k_1^2+\vec k_2^2)}{2M^2Q^2}
-\frac{(\vec k_1^2-\vec k^2_2)^2}{M^2Q^2}\(\frac{1}{2}-\frac{(\vec k_1 \vec k_2)}{\tau^2}\),h^\perp_1h^\perp_1]\Big\},
\\\label{def:S1}
\Sigma_1&=&\frac{4\pi \alpha_{\text{em}}^2}{3N_cs}\sum_{q,G,G'} Q^4\Delta_G^*\Delta_{G'}\Big\{
z_{+\ell}^{GG'}z_{+q}^{GG'}\mathcal{C}[\frac{(\vec  t\vec k_1)-(\vec t\vec k_2)}{Q}\frac{\sqrt{\lambda(\vec k_1^2,\vec k_2^2,\tau^2)}}{\tau^2},f_1f_1]
\\\nn &&
+z_{+\ell}^{GG'}r_{+q}^{GG'}\mathcal{C}[
\frac{(\vec t\vec k_2)-(\vec  t\vec k_1)}{Q}\frac{(\vec k_1\vec k_2)}{M^2}\frac{\sqrt{\lambda(\vec k_1^2,\vec k_2^2,\tau^2)}}{\tau^2}
,h^\perp_1h^\perp_1]\Big\}
\\\nn
\Sigma_2&=&\frac{4\pi \alpha_{\text{em}}^2}{3N_cs}\sum_{q,G,G'} Q^4\Delta_G^*\Delta_{G'}\Big\{
z_{+\ell}^{GG'}z_{+q}^{GG'}\mathcal{C}[\frac{2((\vec  t\vec k_1)-(\vec t\vec k_2))^2-(\vec k_1-\vec k_2)^2}{Q^2}-\frac{(\vec k_1^2-\vec k_2^2)^2}{Q^2\tau^2},f_1f_1]
\\ && \label{def:S2}
+z_{+\ell}^{GG'}r_{+q}^{GG'}\mathcal{C}[
\frac{\vec k_1^2+\vec k_2^2-((\vec  t\vec k_1)-(\vec t\vec k_2))^2}{M^2} +\frac{(\vec k_1^2-\vec k_2^2)^2(\vec k_1\vec k_2)}{M^2Q^2\tau^2}
\\\nn &&\qquad
+\frac{(2((\vec  t\vec k_1)-(\vec t\vec k_2))^2-(\vec k_1-\vec k_2)^2)(\vec k_1^2+\vec k_2^2)-(\vec k_1^2-\vec k_2^2)^2}{2M^2Q^2}
,h^\perp_1h^\perp_1]\Big\},
\\\label{def:S3}
\Sigma_3&=&\frac{4\pi \alpha_{\text{em}}^2}{3N_cs}\sum_{q,G,G'} Q^4\Delta_G^*\Delta_{G'}
z_{-\ell}^{GG'}z_{-q}^{GG'}\mathcal{C}[2\frac{(\vec  t\vec k_1)-(\vec t\vec k_2)}{\tau},\{f_1f_1\}_A]
,
\\\label{def:S4}
\Sigma_4&=&\frac{4\pi \alpha_{\text{em}}^2}{3N_cs}\sum_{q,G,G'} Q^4\Delta_G^*\Delta_{G'}
z_{-\ell}^{GG'}z_{-q}^{GG'}\mathcal{C}[2\frac{\sqrt{\lambda(\vec k_1^2,\vec k_2^2,\tau^2)}}{\tau Q},\{f_1f_1\}_A]
,
\\\label{def:S5}
\Sigma_5&=&\frac{4\pi \alpha_{\text{em}}^2}{3N_cs}\sum_{q,G,G'} Q^4\Delta_G^*\Delta_{G'}
i z_{+\ell}^{GG'}r_{-q}^{GG'}\mathcal{C}[\frac{2(\vec  t\vec k_1)(\vec t\vec k_2)-(\vec k_1\vec k_2)}{M^2}\frac{\sqrt{\lambda(\vec k_1^2,\vec k_2^2,\tau^2)}}{\tau Q} ,\{h^\perp_1h^\perp_1\}_A],
\\\label{def:S6}
\Sigma_6&=&\frac{4\pi \alpha_{\text{em}}^2}{3N_cs}\sum_{q,G,G'} Q^4\Delta_G^*\Delta_{G'}
i z_{+\ell}^{GG'}r_{-q}^{GG'}
\\\nn &&\qquad\qquad
\mathcal{C}[\frac{(\vec  t\vec k_1)-(\vec t\vec k_2)}{\tau M^2}\(-(\vec k_1\vec k_2)+2\frac{\vec k_1^2\vec k_2^2-(\vec k_1\vec k_2)^2}{Q^2}\),\{h^\perp_1h^\perp_1\}_A],
\\\label{def:S7}
\Sigma_7&=&0.
\end{eqnarray}
Here, $\vec t=\vec q_T/|\vec q_T|$ and $\lambda$ is the triangle function, i.e.,
\begin{eqnarray}
\lambda(a,b,c)=a^2+b^2+c^2-2ab-2ac-2bc.
\end{eqnarray}
The coupling constants $z_{\pm q}^{GG'}$ are defined in (\ref{def:z}), and the subscript $q$ indicates that they are taken with the quark $q$ quantum numbers. Meanwhile, the coupling constants $r_{\pm q}^{GG'}$ read 
\begin{eqnarray}
r^{GG'}_{+q}&=&2(g^R_Gg^L_{G'}+g^L_Gg^R_{G'})=\frac{v^Gv^{G'}-a^{G}a^{G'}}{4s_W^2c_W^2},
\\
r^{GG'}_{-q}&=&2(g^R_Gg^L_{G'}-g^L_Gg^R_{G'})=-\frac{v^Ga^{G'}-a^{G}v^{G'}}{4s_W^2c_W^2}.
\end{eqnarray}
All the information related to the EW coupling constants is summarized in the appendix \ref{app:couplings}. Finally, $\mathcal{C}$ denotes the convolution, which is defined as
\begin{eqnarray}\label{def:convolution} 
\mathcal{C}[A,f_1f_2]&=&4p_1^+p_2^-C_0\(\frac{Q^2}{\mu^2}\)\int d\xi_1 d\xi_2 \int d^4k_1 d^4k_2 
\delta^{(4)}(q-k_1-k_2)\delta(k_1^2)\delta(k_2^2)
\\\nn &&
\delta(k_1^+-\xi_1p_1^+)\delta(k_2^--\xi_2p_2^-)
A \(f_{q1}(\xi_1,k_{1T}^2)f_{\bar q2}(\xi_2,k_{2T}^2)+f_{\bar q1}(\xi_1,k_{1T}^2)f_{q2}(\xi_2,k_{2T}^2)\),
\end{eqnarray}
where $f_q$ and $f_{\bar q}$ are the quark and anti-quark distributions, correspondingly. Furthermore, the convolutions with the second argument labelled as $\{f_1f_2\}_A$ should be taken with an anti-symmetric combination of the quark and anti-quark distributions, i.e.,
\begin{eqnarray}
\mathcal{C}[A,\{f_1f_2\}_A]&=&4p_1^+p_2^-C_0\(\frac{Q^2}{\mu^2}\)\int d\xi_1 d\xi_2 \int d^4k_1 d^4k_2 
\delta^{(4)}(q-k_1-k_2)\delta(k_1^2)\delta(k_2^2)
\\\nn &&
\delta(k_1^+-\xi_1p_1^+)\delta(k_2^--\xi_2p_2^-)
A \(f_{q1}(\xi_1,k_{1T}^2)f_{\bar q2}(\xi_2,k_{2T}^2)-f_{\bar q1}(\xi_1,k_{1T}^2)f_{q2}(\xi_2,k_{2T}^2)\).
\end{eqnarray}
In these formulas we have omitted the TMD distributions scaling argument, which is $(\mu,Q^2)$ for all of them. A synopsis of some of the $\Sigma_n$'s properties is presented in table \ref{tab:sigmas}.

During the derivation of the angular coefficients we have eliminated all the contributions linear in $(\vec k_1\times \vec k_2)$ because they vanish in the convolution with functions that depend only on scalar products (see appendix \ref{app:convolution}). Such terms appear in $\Sigma_{0,1,2,5,6,7}$. Besides, the structure function $\Sigma_7$ contains only one term, which is proportional to $(\vec k_1\times \vec k_2)$, and is thus completely eliminated. This indicates that $\Sigma_7$ is a pure higher-twist function from the perspective of the TMD factorization.

The expressions (\ref{def:SU}-\ref{def:S7}) constitute the main theoretical result of the paper. They are novel, and only the LP terms can be compared with the literature. It is worth mentioning that the convolution integrands (\ref{def:S0}-\ref{def:S6}) can be presented in various alternative forms using kinematic relations. In the current exposition, we have ordered the expressions by powers, and made the symmetries with respect to $\vec k_1\leftrightarrow\vec k_2$ explicit. Practically, it is more convenient to use proper constructed variables that simplify the integration. These variables, as well as the corresponding expressions for the integrands, are presented in appendix \ref{app:convolution}.

The fiducial cross-section can be written in terms of the structure functions $\Sigma_n$. It reads
\begin{eqnarray}
\frac{d\sigma_{\text{fid}}}{d^4q}=\sum_{n=U,0,...,7}\mathcal{P}_n(\text{cuts})\Sigma_n,
\end{eqnarray}
where the cut factors are defined in eqn.~(\ref{P-general:app}). Note that this formula is universal and does not rely on the factorization properties of the $\Sigma_n$'s. Thus, it can be used together with eqn.~(\ref{def:sigma_dOmega}) for cross-comparing systematic uncertainties in the fiducial and angular cross-sections measurements.

The photon does not couple to the axial current, so the EW couplings $z_-^{GG'}$ and $r_-^{GG'}$ are zero for the photon channel. Therefore, only the structure functions $\Sigma_{U,0,1,2}$ contribute to the photon production, while the rest emerge purely due to the weak current. The structures $\Sigma_{5,6}$ (and the vanished $\Sigma_7$) are proportional to $r_{-q}^{GG'}$, which is anti-symmetric under $G\leftrightarrow G'$. As a result, it appears only due to the imaginary part of the interference between $\gamma$ and $Z$ bosons, and has an extra suppression factor of $\sim \Gamma_Z/M_Z$. 

In this work, we mainly study the neutral-boson current; however, it is simple to generalize the expressions for the case of the charged current. One has to take into account that the sum over the flavors is not diagonal but mixed via the CKM-matrix. It should be noted that, because of the absence of the right coupling for the W-boson, $g_R^W=0$ and the couplings $r_\pm^{WW}=0$. Consequently, all contributions proportional to the Boer-Mulder function vanish. This results in a different leading behavior for the neutral and charged currents structure functions.

The leading power expression is given by a leading term of $Q^2\to 0$, keeping other scales fixed. The variables $\vec k_{1,2}^2$ are integrated in the region which reaches values of $(\tau+|\vec q_T|)^2/4\sim Q^2/4$, and thus cannot be formally considered small. Nonetheless, the restoration of the LP expression is obtained if the transverse scales $\vec q_T$, $\vec k_1$ and $\vec k_2$ are considered fixed. In this limit one gets
\begin{eqnarray}\label{large-Q}
\lim_{Q^2\to \infty}\mathcal{C}[A,f_1f_2]&=&\frac{2}{Q^2}\int d^2\vec k_1 d^2\vec k_2
\delta^{(2)}(\vec q_T-\vec k_1-\vec k_2)
\\\nn &&
(f_{1q}(x_1,\vec k_{1}^2)f_{2\bar q}(x_2,\vec k_{2}^2)
+f_{1\bar q}(x_1,\vec k_{1}^2)f_{2q}(x_2,\vec k_{2}^2))
\lim_{Q^2\to \infty}A.
\end{eqnarray}
Only four structure functions $\Sigma_{U,2,4,5}$ have LP contribution. Among them, the function $\Sigma_{2}$ is well known. It is often referred to as the $F_{UU}^{\cos2\phi}$ structure function, and it has been discussed in many articles (see, for instance, \cite{Barone:2010gk, Lu:2011mz, Liu:2012fha, Pasquini:2014ppa, Wang:2018naw, Bacchetta:2019qkv}). The LP structure functions $\Sigma_{4,5}$ are less known. Some discussion on them can be found in ref.~\cite{Scimemi:2019cmh, Ebert:2020dfc}. The structure functions $\Sigma_{1,3,6}$ are $\sim Q^{-1}$, and $\Sigma_{0}$ is $\sim Q^{-2}$. Note that, aside from their power suppression with respect to $Q$, the structure functions $\Sigma_{1,3,6,7}$ vanish in the limit $q_T\to0$. The remaining structure functions have a non-zero contribution at $q_T=0$.

\begin{table}
\begin{center}
\renewcommand{\arraystretch}{2.5}
\begin{tabular}{|c|c|c|c|c|}\hline
 & Leading behavior & \thead{EW \\channel} & \thead{$y$ \\symmetric} & \thead{Collinear \\ factorization}
\\\hline\hline
$\Sigma_U$ & $f_1f_1$ & all & S & $\delta(q_T)+\mathcal{O}(\alpha_s)$
\\\hline
$\Sigma_0$ & $\Ds\frac{\vec \Delta^2}{Q^2}\Big(f_1f_1+\frac{\vec \Delta^2}{4M^2}h_1^\perp h_1^\perp\Big)$ & all & S  &  $\mathcal{O}(\alpha_s)$
\\\hline
$\Sigma_1$ & $\Ds\frac{(\vec t\vec \Delta)}{Q}\Big(f_1f_1+\frac{\vec \Delta^2}{4M^2}h_1^\perp h_1^\perp\Big)$ & all & A  & $\mathcal{O}(\alpha_s)$
\\\hline
$\Sigma_2$ & $\Ds \frac{\vec \Delta^2-2(\vec t\vec \Delta)^2}{2M^2}\Big(h_1^\perp h_1^\perp-\frac{2M^2}{Q^2}f_1f_1\Big)$ & all & S  & $\mathcal{O}(\alpha_s)$
\\\hline
$\Sigma_3$ & $\Ds 2\frac{(\vec t\vec \Delta)}{Q}\{f_1f_1\}_A$ & $ZZ$, $WW$ & S & $\mathcal{O}(\alpha_s)$
\\\hline
$\Sigma_4$ & $\Ds 2\{f_1f_1\}_A$ & $ZZ$, $WW$ & A & $\delta(q_T)+\mathcal{O}(\alpha_s)$
\\\hline
$\Sigma_5$ & $\Ds \frac{\vec \Delta^2-2(\vec t\vec \Delta)^2}{4M^2}\{h^\perp_1h^\perp_1\}_A \times \Big[\frac{\Gamma_Z}{Q}\Big]$ & $Z\gamma$ & A & $\mathcal{O}(\alpha_s^2)$
\\\hline
$\Sigma_6$ & $\Ds \frac{(\vec t\vec \Delta)}{Q}\frac{\vec \Delta^2}{4M^2}\{h_1^\perp h_1^\perp\}_A \times \Big[\frac{\Gamma_Z}{Q}\Big] $ & $Z\gamma$ & S & $\mathcal{O}(\alpha_s^2)$
\\\hline
$\Sigma_7$ & $\Ds 0$ & $Z\gamma$ & & $\mathcal{O}(\alpha_s^2)$ \\\hline
\end{tabular}
\caption{\label{tab:sigmas} The synopsis of the main properties of the structure functions $\Sigma_n$. The column ``leading behavior'' shows the term that survives in the limit $Q\to\infty$ (\ref{large-Q}) and $q_T\to0$, with $\vec \Delta=\vec k_1-\vec k_2$ and $\vec t=\vec q_T/|\vec q_T|$. For $\Sigma_{5,6}$ we also indicate the extra suppression factor coming from the lepton tensor. The column ``y-symmetric'' indicates whether the structure function is symmetric (S) or anti-symmetric (A) with respect to the $y\to-y$ transformation. The column ``Collinear factorization'' marks the order at which the collinear factorization produces a non-zero contribution to the structure function \cite{Mirkes:1994eb}.}
\end{center}
\end{table}

\section{Perturbative and non-perturbative setup}
\label{sec:setup}

The TMD factorization theorem with KPCs is a generalization of the ordinary TMD factorization. Therefore, it might be expected to describe the data sufficiently well without significant modifications of the models for TMD distributions. Indeed, we have found a good agreement between the theoretical predictions of the new expressions and the standard pool of DY data using the TMDPDFs from the ART23 extraction \cite{Moos:2023yfa}. However, the value of $\chi^2$ per point (for 627 points) increased from $1.04$ (in ART23 at LP) to $1.8$. This happens mainly due to the very precise data taken at ATLAS \cite{ATLAS:2015iiu, ATLAS:2019zci}, CMS \cite{CMS:2019raw}, and LHCb \cite{LHCb:2021huf}, where even a minor change leads to a huge increase of $\chi^2$. In the long run, it indicates that one should perform a new global fit of these data with the updated formula. Performing an accurate global fit is a complex task which goes beyond the scope of the present article. 

In this work, as a first exploratory study, we use the input of ART23 (with the latest update \cite{Moos:INPREP}), and perform a tuning of the non-perturbative parameters for the central value. The modification of the non-perturbative parameters that we found is not large, and it does not change the central value significantly. Nonetheless, it helps to describe the shape of the highly precise LHC measurements better. In contrast, the Boer-Mulders function is unknown, and we estimate it using the data for the $A_2$ angular distribution. In this section, we describe the main points of the ART23 setup (for all details we refer to the original work \cite{Moos:2023yfa}), describe the changes made here, and discuss the impact of the inclusion of KPCs on the theory predictions.

\subsection{Perturbative input and evolution}

The factorization theorem with KPCs inherits all the perturbative properties of the LP factorization theorem. It implies the same hard coefficient function and the same evolution. All these ingredients of TMD factorization are known up to a high-loop order. Still, there is a significant technical difference in the realization of the LP factorization expression and the improved one. Namely, the LP factorization is formulated in the position space, while the summation of KPCs is performed in the momentum space. 

The position-space TMD distributions are related to the momentum-space distributions by the Fourier/Hankel transform. For the unpolarized and Boer-Mulders functions the transformation is defined as \cite{Boer:2011xd}
\begin{eqnarray}
\tilde f_1(x,b;\mu,\zeta)&=&2\pi \int_0^\infty dk_T k_T J_0(bk_T) f_1(x,k_T;\mu,\zeta),
\\
\tilde h_1^\perp(x,b;\mu,\zeta)&=&\frac{2\pi}{M^2} \int_0^\infty dk_T \frac{k^2_T}{b} J_1(bk_T) h_1^\perp(x,k_T;\mu,\zeta),
\end{eqnarray}
where $J_n$ is the Bessel function of the first kind. The evolution of the TMD distributions is given by the following pair of equations
\begin{eqnarray}\label{def:evol}
\frac{d}{d\ln\mu^2}\ln F(x,b;\mu,\zeta)=\frac{\Gamma_{\text{cusp}}(\mu)}{2}\ln\(\frac{\mu^2}{\zeta}\)-\frac{\gamma_V(\mu)}{2},
\qquad
\frac{d}{d\ln\zeta}\ln F(x,b;\mu,\zeta)=-\mathcal{D}(b,\mu),
\end{eqnarray}
where $F$ is any TMD distribution of twist-two (including $f_1$ and $h_1^\perp $), $\Gamma_{\text{cusp}}$ is the cusp anomalous dimension, $\gamma_V$ is the light-like-quark anomalous dimension, and $\mathcal{D}$ is the Collins-Soper kernel. 

The code \texttt{artemide} is based on the $\zeta$-prescription realization of the TMD evolution \cite{Scimemi:2018xaf, Vladimirov:2019bfa}. In this case, the defining scale for TMD modeling is $\zeta_\mu(b)$, which is defined non-perturbatively such that the optimal TMD distribution $F(x,b)=F(x,b;\mu,\zeta_\mu(b))$ is $\mu$-independent. The optimal TMD distribution is then evolved to a desired scale along a $\mu=\,\,$const. path, and then transformed to the momentum space. In this way, our expressions for the TMD distributions are
\begin{eqnarray}
f_1(x,k_T;\mu,\zeta)&=&\int \frac{db}{2\pi}bJ_0(bk_T)\(\frac{\zeta}{\zeta_\mu(b)}\)^{-\mathcal{D}(b,\mu)}\tilde f_1(x,b),
\\
h_1^\perp(x,k_T;\mu,\zeta)&=&\frac{M^2}{k_T}\int \frac{db}{2\pi}b^2J_1(bk_T)\(\frac{\zeta}{\zeta_\mu(b)}\)^{-\mathcal{D}(b,\mu)}\tilde h^\perp_1(x,b),
\end{eqnarray}
where $\tilde f_1(x,b)$ and $\tilde h^\perp_1(x,b)$ are the optimal unpolarized and Boer-Mulders distributions.

The Collins-Soper kernel is a non-perturbative function that describes the QCD vacuum \cite{Vladimirov:2020umg}. The ART23 model for the Collins-Soper kernel reads
\begin{eqnarray}\label{CS-kernel}
\mathcal{D}(b,\mu)=\mathcal{D}_{\text{small-b}}(b^*,\mu^*)+\int \frac{d\mu'}{\mu'}\Gamma_{\text{cusp}}+bb^*\(c_0+c_1\ln\(\frac{b^*}{B_{\text{NP}}}\)\),
\end{eqnarray}
where $\mathcal{D}_{\text{small-b}}$ is the perturbative small-$b$ part \cite{Echevarria:2015uaa, Vladimirov:2017ksc}, $B_{\text{NP}}=1.5$GeV$^{-1}$ (fixed), $c_{0,1}$ are free parameters, and
\begin{eqnarray}
b^*=\frac{b}{\sqrt{1+b^2/B^2_\text{NP}}},\qquad \mu^*(b)=\frac{2 e^{-\gamma_E}}{b^*}.
\end{eqnarray}
This model obey all the required properties of the Collins-Soper kernel and describes the data well. The parameters $c_{0,1}$ were determined in the ART23 fit.

In this work we use the same perturbative setup as in ART23. It can be generally qualified as N$^4$LL. It includes: the hard coefficient function $C_0$ at N$^4$LO ($\sim \alpha_s^4$) \cite{Lee:2022nhh}, the light-like-quark anomalous $\gamma_V$ at N$^3$LO ($\sim\alpha_s^4$) \cite{Agarwal:2021zft} and the small-$b$ matching for the Collins-Soper kernel at N$^3$LO ($\sim \alpha_s^4$) \cite{Duhr:2022yyp, Moult:2022xzt}. The cusp anomalous dimension $\Gamma_{\text{cusp}}$ is taken at N$^4$LO order ($\sim \alpha_s^5$, central value) \cite{Herzog:2018kwj}. The small-$b$ matching for the unpolarized distributions (used in the modeling of $f_1$) is at N$^3$LO order ($\sim \alpha_s^3$) \cite{Echevarria:2016scs, Luo:2020epw}. As the collinear input we use the MSHT20 unpolarized PDF \cite{Bailey:2020ooq}, which is extracted at N$^2$LO order ($\sim\alpha_s^3$) (this is the only part that deviates from the strict definition of N$^4$LL order, that is why in some literature this setup is referred to as N$^{4-}$LL).

The numerical evaluation of the factorization with KPCs is much slower than that of the LP factorization. The reason is that an independent Hankel transform must be performed for each call of the integrand in the convolution. For comparison, the LP factorization requires only a single call of the Hankel transform. Moreover, the values of $k_T$ within the convolution integral can be large, which sets a serious restriction for the algorithm. For example, the Ogata quadrature algorithm \cite{Ogata:2005}, which is used by many codes, is very efficient for small values of $k_T$, but its efficiency drops rapidly at large $k_T$. Meanwhile, the convolution $\mathcal{C}$ requires the values of TMD up to $|\vec k_T|=(\tau+|\vec q_T|)/2$, which reaches $\sim 100$GeV for Z-boson production measurements, and even higher for high-energy Drell-Yan measurements \cite{CMS:2022ubq}. Therefore, in the updated version of \texttt{artemide} we have implemented the algorithm based on the Levin system of differential equations \cite{Levin:1982, Levin:1996}. The efficiency of this approach does not drop significantly at large $q_T$, and it can be formulated as a grid transformation of a grid in $b$-space to a grid in $q_T$-space by a matrix multiplication. In many aspects our algorithm is analogous to the one recently presented in ref.~\cite{Diehl:2024mmc}. Details of the implementation are presented in appendix \ref{app:Fourier}.

The implementation of new algorithms leads to a significant improvement of the computational speed of \texttt{artemide} (also for the LP part); however, the computation time of the TMD-with-KPCs expression is much longer than the analogous computation at the LP approximation. This is the main limitation that prevents us from performing an accurate fit of the distributions. Such fit is possible but requires extra technical preparations and computer power.

\subsection{Non-perturbative model for $f_1$}
\label{sec:NP-setup}

The ART23 fit utilizes the following model for the optimal unpolarized TMD distribution $f_1$
\begin{eqnarray}
\tilde f_{1q}(x,b)=\int_x^1\frac{dy}{y}\sum_{q'}C_{q\ot q'}(y,b,\mu_{\text{OPE}})q_{q'}\(\frac{x}{y},\mu_{\text{OPE}}\)f^q_{\text{NP}}(x,b),
\end{eqnarray}
where $C_{q\to q'}$ is the coefficient of the small-$b$ matching (used at N$^3$LO \cite{Luo:2020epw}) and $q(x,\mu)$ is the collinear PDF (we are using MSHT20 extraction \cite{Bailey:2020ooq}). The operator product expansion is done at the scale
\begin{eqnarray}\label{muOPE}
\mu_{\text{OPE}}=\frac{2e^{-\gamma_E}}{b}+5\text{GeV},
\end{eqnarray}
where the 5GeV constant is added to guarantee that $\mu$ does not approach the Landau pole at large values of $b$. The function $f_{\text{NP}}^q$ is a non-perturbative ansatz that is taken in the form
\begin{eqnarray}
f_{\text{NP}}^q(x,b)=\frac{1}{\cosh\(\(\lambda_1^q(1-x)+\lambda_2^q x\)b\)},
\end{eqnarray}
where $\lambda_{1,2}^q$ are free parameters. In ART23 $u$, $d$, $\bar u$, $\bar d$ and $sea$ flavors are considered. The inclusion of the separate parameters for flavors is important, since it helps to better describe the data \cite{Bacchetta:2024qre} and overcome the problem of PDF-bias \cite{Bury:2022czx}.

The parameters $\lambda_{1,2}^q$ were fit in ART23 to a large set of DY data, together with the parameters $c_{0,1}$ of the Collins-Soper kernel. Their values can be found in the ART23 publication \cite{Moos:2023yfa}. In this work, we use the latest update of this extraction, which will be released soon \cite{Moos:INPREP}. The differences from ART23 are the following
\begin{itemize}
\item The parameter $B_{\text{NP}}$ in (\ref{CS-kernel}) is fixed $B_{\text{NP}}=1.5$GeV$^{-1}$, instead of fitting (the fitted value $B_{\text{NP}}=1.56^{+0.13}_{
-0.09}$GeV$^{-1}$).
\item The constant parameter in (\ref{muOPE}) was fixed to 5GeV (instead of 2GeV). This allows to avoid the problem of the quark-mass-thresholds in the coefficient function.
\item We fix $\lambda_1^u=\lambda_1^{\bar u}$ and $\lambda_1^d=\lambda_1^{\bar d}$. This restriction is required to support better the properties of the valence TMD distribution $f_{\text{val}}(x,b)=f_{q}(x,b)-f_{\bar q}(x,b)$. Namely, with $\lambda_1^q=\lambda^{\bar q}_1$ one has a finite integral $\int_0 dx f_{\text{val}}(x,b)$ for all values of $b$. This is helpful in many aspects. In particular, it leads to a well-behaving second transverse-momentum moment \cite{delRio:2024vvq}.
\end{itemize}
These modifications slightly reduce the quality of the LP fit, but make it more stable. 

The ART23 extraction is tuned to 627 data points collected at low (FermiLab and RHIC experiments \cite{Ito:1980ev, Moreno:1990sf, E772:1994cpf, PHENIX:2018dwt}) and high (Tevatron and LHC \cite{CDF:1999bpw, CDF:2012brb, D0:1999jba, D0:2007lmg, D0:2010dbl, ATLAS:2015iiu, ATLAS:2019zci, CMS:2011wyd, CMS:2016mwa, CMS:2019raw, CMS:2022ubq, LHCb:2015mad, LHCb:2021huf}) energies. The data span in $Q$ from 4 to 1000 GeV, and in $x$ down to $\sim x^{-4}$. The details of the description of each experiment and the constriction of the $\chi^2$-test function can be found in ref.~\cite{Moos:2023yfa}.

The original ART23 fit has $\chi^2/N_{pt}=0.96$. After the modification described above it has $\chi^2/N_{pt}=1.04$ with a better distribution of parameters and $\chi^2$ between experiments. Using it as input for the KPCs formula we set $\chi^2/N_{pt}\sim 1.8$. The worst $\chi^2$ is obtained for the the very precise Z-boson measurements by ATLAS, CMS and LHCb. In these cases, even a 1\% modification in the shape leads to a significant increase of $\chi^2$. In the aforementioned measurements, the changes induced by the new formula are quite significant due to the different cut-factor. ART23 uses the LP cut factor, which has somewhat a different shape in $q_T$ (see sec.~\ref{sec:sigmaU}). To find a better agreement with the data, we made the central value fit of the parameters using the ART23 model. We have found the following values
\begin{align}\nn
&\lambda_1^u=0.42, && \lambda_2^u=0.17, && \lambda_2^{\bar u}=5.9,
\\
&\lambda_1^d=0.56, && \lambda_2^d=8.4, && \lambda_2^{\bar d}=5.6,
\\\nn
&\lambda_1^{\text{sea}}=1.1, && \lambda_2^{\text{sea}}=0.1. 
\end{align}
At this values of $\lambda$ we have $\chi^2/N_{pt}=1.2$, which is not perfect, but satisfactory for the first study. We found that the majority of parameters remain within 1$\sigma$ from the ART23 fit, except $\lambda_2^{u,s,\text{sea}}$.

The computation of the new uncertainty band requires greater computer power. Therefore, we utilize the uncertainty band of ART23 and shift it to a new central value. We expect this procedure to provide a reliable estimation because the central value is not significantly modified, and the data set that determines the uncertainty band is the same. Note that the uncertainty band of ART23 incorporates uncertainties from the data, non-perturbative parameters and the PDF uncertainty.

\subsection{Non-perturbative model for $h_1^\perp$}
\label{sec:h1}

There have been a few attempts to extract the Boer-Mulders function $h_1^\perp$ from the data \cite{Barone:2009hw, Barone:2010gk, Barone:2015ksa, Bastami:2018xqd}. In these studies, the main source of the information about the Boer-Mulders function is Semi-Inclusive Deep-Inelastic scattering (SIDIS), in which the Boer-Mulders function is multiplied by the Collins function (which must also be determined from the data). The second source is the E866 experiment at NuSea \cite{NuSea:2008ndg}, which is problematic for TMD factorization (see discussion in sec.~\ref{sec:nu}). These studies are done at the leading perturbative order. In this way, these estimations of the Boer-Mulders function could not be used in our work.

In this study, we define the Boer-Mulders function from the comparison with the data for the $A_2$ angular distribution measured at ATLAS. In this case, it is crucial to take into account the effects of evolution. The $\zeta$-prescription exactly separates the effects of the evolution from the optimal distribution. Therefore, one can use the evolution at any perturbative order without the need to match it with the small-$b$ coefficient function (as required in the Collins-Soper-Sterman approach \cite{Collins:2011zzd, Aybat:2011zv}). As a consequence, we can use our N$^4$LL setup with a purely non-perturbative optimal Boer-Mulders function. This is a well-established approach that has been used, for example, in refs.~\cite{Bury:2020vhj, Bury:2021sue, Horstmann:2022xkk} to simultaneously determine the Sivers and worm-gear-T functions from high- and low-energy data.

We model the optimal Boer-Mulders with the following function
\begin{eqnarray}\label{h1:model}
\tilde h_1^\perp(x,b)=\frac{2^{\alpha+1}}{\Gamma(\alpha+1)}\frac{N}{\cosh( \lambda b)}\,x\ln^\alpha(1/x).
\end{eqnarray}
This functional form is motivated by the subsequent observations:
\begin{itemize}
\item At $b\to0$ the Boer-Mulders function turns to $\pi E(-x,0,x;\mu)$ \cite{Scimemi:2018mmi, Rein:2022odl}, which grows at small-x and at large-$\mu$ (and hence at small-$b$) due to the effects of evolution \cite{Rodini:2024usc}. This is simulated by $\ln^\alpha(1/x)$.
\item Still, at $x\to 0$ the function $h_1^\perp$ behaves $\sim x$ as it is suggested in ref.~\cite{Kovchegov:2022kyy}. 
\item At $x\to1$ the function turns to zero as $(1-x)^\alpha$.
\item The $b-$profile is taken similar to ART23.
\item The expression is normalized to the integral over $x$, in order to decorrelate parameters $\alpha$ and $N$.
\end{itemize}
We consider the same shape for all flavors, since there is no data sensitive to the flavor differences. The DY reaction is proportional to the product $h_{1q}^\perp h_{1\bar q}^\perp$, and thus cannot determine the sign of the Boer-Mulders function, but only the relative sign between the quark and anti-quark distributions. To resolve this sign we use $|N|$ for quark distributions and $N$ for anti-quark distributions. In other words, the positive(negative) $N$ indicates the same(opposite) sign between the quark and anti-quark distributions.

\begin{figure}
\begin{center}
\includegraphics[width=0.45\textwidth]{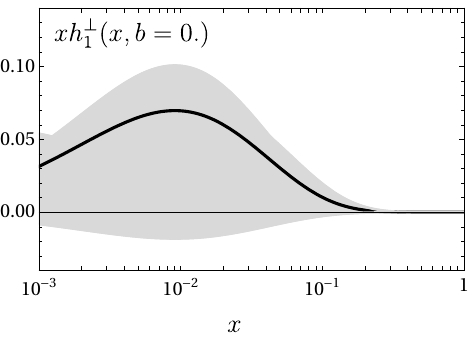}
\end{center}
\caption{\label{fig:h1} Optimal Boer-Mulders function at $b=0$, determined from the ATLAS data.}
\end{figure}

To fix the parameters, we employ the ATLAS measurement of the angular distribution $A_2$ \cite{ATLAS:2016rnf}. We consider data points with $q_T<10$GeV to be safely within the TMD-factorization region. ATLAS provides measurements in four bins in $|y|$ ($[0,1]$, $[1.,2.]$, $[2.,3.5]$) and in $2.5$GeV-wide bins in $q_T$. Thus, we have in total $N_{pt}=12$ data points to restrict our ansatz. It is clearly insufficient, and so we set the parameter $\lambda=0.2$GeV, since the LHC data is not very sensitive to large-$b$ behavior. Afterwards, we perform a fit of the parameters $\alpha$ and $N$. The fit converges to $\chi^2/N_{pt}=1.16$. To estimate the uncertainty, we vary $N$ and $\alpha$ independently (keeping another parameter fixed) and find the boundary $\chi^2/N_{\text{pt}}=2$, which corresponds to 98\%CI for 12 data points. Additionally, we demand that $\chi^2_{\text{NuSea}}/N_{pt}<3.$ (for the discussions on issues with the NuSea data see sec.~\ref{sec:nu}). The latter requirement restricts the parameter $\alpha$ from above. This procedure gives us a rough estimation of the uncertainty band for the Boer-Mulders function. We obtain the following values for the parameters
\begin{eqnarray}
\lambda=0.2\text{GeV}(\text{fixed}),\qquad N=-0.27_{-0.12}^{+0.34},\qquad\alpha=9.4_{-0.9}^{+5.4}.
\end{eqnarray}
Varying the parameter $\lambda$ by a reasonable amount does not affect $\chi^2$. The procedure we used is not precise, but it is sufficient given the poor quality of the present data set. The plot of the obtained Boer-Mulders function is shown in fig.\ref{fig:h1}.

Our rough estimation of the Boer-Mulders function clearly prefers $N<0$, i.e. the relative sign of the quark and anti-quark distributions is preferably negative. This is required by the positivity of $A_2$ observed at ATLAS (and also CMS, LHCb and CDF). In sec.~\ref{sec:nu} we demonstrate that it is also in a general agreement with the low-energy data. The negative relative sign contradicts the extraction in ref.~\cite{Barone:2009hw, Barone:2010gk}, where the analysis was done in a much simpler framework.

It is also important to mention that the equality of the Boer-Mulders function for quarks and anti-quarks used by us results into the zero $A_5$ and $A_6$ distributions. The only measurement of these structure functions is provided by ATLAS \cite{ATLAS:2016rnf}. Both angular distributions are measured to be very small $<0.5\%$ (in the TMD factorization region) with significant uncertainties that are generally larger than the measured value. The same is true for the angular distribution $A_7$. Therefore, the current prediction $A_{5,6,7}=0$ does not contradict the measurement. 

\section{Description of angular structure functions}
\label{sec:practice}

In this section, we present the comparison of the predictions by the TMD-with-KPCs factorization approach with the angular distributions of the unpolarized DY reaction. We start with the integrated cross-section, which, although is not the subject of this study, serves as the normalization for all angular coefficients. Then, we proceed with the angular coefficients, ordered by their (subjective) importance for TMD physics. It should be noted that we do not discuss the functions $A_{5,6,7}$ since the current theoretical prediction $A_{5,6,7}=0$ is in agreement with the data.

The main source of data for the angular distributions $A_n$ is the measurement performed by ATLAS \cite{ATLAS:2016rnf}, made at $\sqrt{s}=8$TeV in the vicinity of the Z-boson peak $80<Q<100$GeV, in 3 bins of $y$ and 2.5GeV-wide bins of $q_T$. For our studies, we have selected data with $q_T<20$GeV, which lies within the TMD-factorization region (although for $q_T>10$GeV one sees a $\sim2-4\%$ contribution of $q_T/Q$ corrections). Besides the ATLAS measurement, there are others from CMS \cite{CMS:2015cyj}, LHCb \cite{LHCb:2022tbc} and CDF \cite{CDF:2011ksg} experiments, but these are only done for some of distributions and in wider $q_T$-bins. There is also a low-energy measurement from NuSea (experiment E866) \cite{NuSea:2008ndg}. However, it is effectively one-dimensional (i.e., integrated over a wide range of kinematic parameters), making it complicated to apply the factorization restrictions reliably. We specify the details of particular measurements in their corresponding sections.

In order to compare with the data one must take into account the bin integration. In our case, we study the distributions $A_n$ defined as $\Sigma_n/\Sigma_U$.  The bin integration is performed separately for the denominator and the numerator, i.e.
\begin{eqnarray}
\langle A_n\rangle_{\text{bin}}=\frac{\langle \Sigma_n\rangle_{\text{bin}}}{\langle \Sigma_U\rangle_{\text{bin}}},
\end{eqnarray}
where
\begin{eqnarray}
\langle \Sigma_n\rangle_{\text{bin}}=\int_{\text{bin}} d^4 q \,\Sigma_n.
\end{eqnarray}
This is consistent with the experimental procedure of first binning the events and then determining the coefficients. It is important to note that the experimental measurements determine the direction of the $z$ axis by the largest component of $q$. This implies that $y$ is always positive, or that the integration measure over $y$  is $\sign(y)$. The uncertainty band for the unpolarized contribution $\sim f_1f_1$ is obtained by computing $\langle A_n\rangle$ for each member of the ART23 replicas' distribution.

\subsection{Angle-integrated cross-section}
\label{sec:sigmaU}

\begin{figure}
\begin{center}
\includegraphics[width=0.32\textwidth]{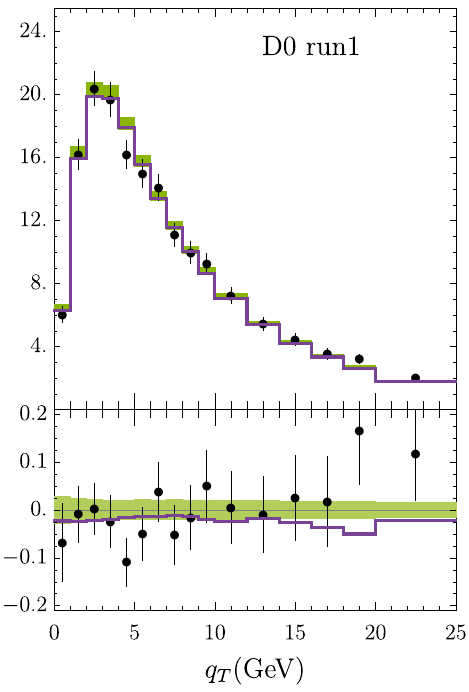}
\includegraphics[width=0.32\textwidth]{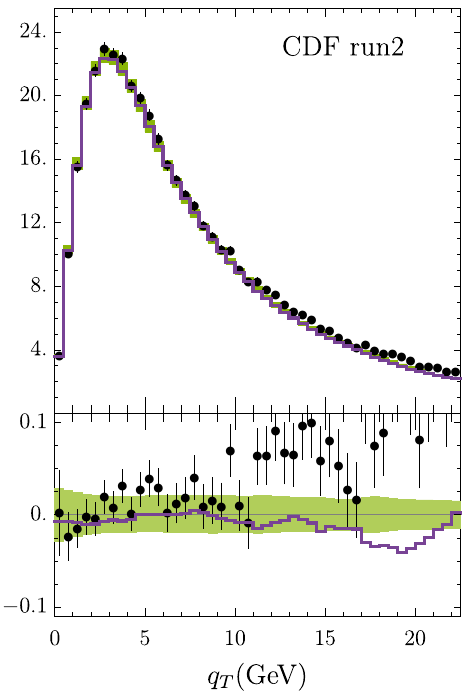}
\end{center}
\caption{\label{fig:unpol-HE} Example of the description of the integrated over angle cross-sections at Tevatron \cite{D0:1999jba, CDF:2012brb}. The band is the uncertainty of the theory prediction taken from ART23. The solid purple line is the LP prediction by ART23.}
\end{figure}

As it is discussed in sec.~\ref{sec:NP-setup}, we have used the ART23 extraction as the baseline for our analysis. The computation of $\chi^2$ for the complete ART23 data set demonstrated an unsatisfactory value of $\chi^2/N_{pt}=1.8$. Therefore, we performed the central value fit, and reduced it to $\chi^2/N_{pt}=1.2$. Simultaneously, we do not modify the model, nor do we include the theory/experimental uncertainty in the fitting procedure. These corrections are not important for a first exploratory study like the one presented here, and we leave a more accurate consideration for the future. 

We have found that the description of the cross-section is satisfactory even without fine tuning. In general, the modified factorization theorem reproduces the shape of the LP cross-section, but increases the total normalization somewhat. At high energies (Tevatron and LHC) the normalization increases by 1-2\%. The example of the high-energy cross-section is given in fig.\ref{fig:unpol-HE}, where we present the comparison with the data collected at the Tevatron in run1 and run2. Clearly, the shape of the theory prediction is almost unchanged (the oscillations visible in the LP predictions are due to the quark thresholds, as discussed in ref.~\cite{Moos:2023yfa}), but the whole curve is shifted up by 1-2\%. 

\begin{figure}
\begin{center}
\includegraphics[width=0.42\textwidth]{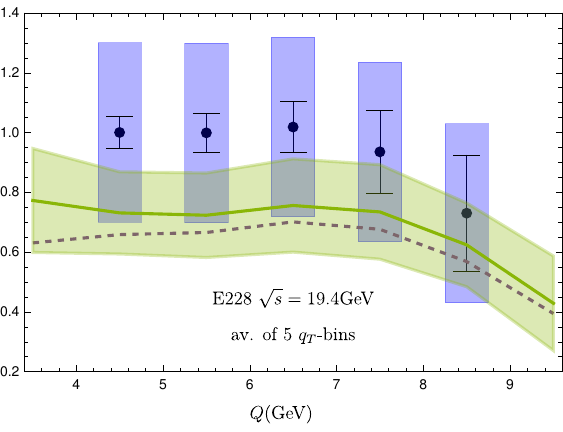}
\includegraphics[width=0.42\textwidth]{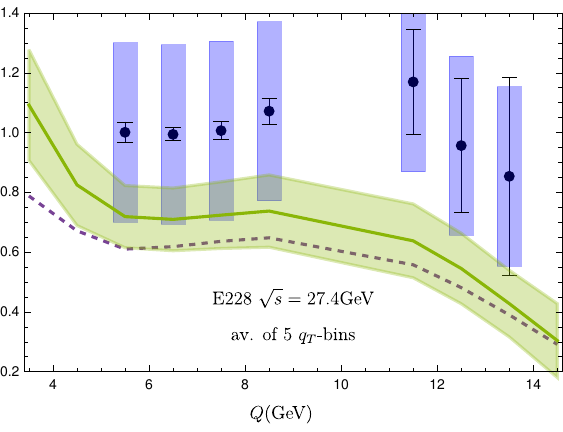}
\end{center}
\caption{\label{fig:unpol-LE} Comparison of the relative normalizations for the E228 experiment \cite{Ito:1980ev} and the theory predictions as a function of $Q$. The normalization is computed by first 5 $q_T$-bins and weighted by an exponential factor $Ne^{-Q}$ (with $N$ tuned to the first experimental point). The error-bands demonstrate the statistical uncertainty, and the blue boxes the luminosity uncertainty. The green line with a band is the prediction of the TMD-with-KPCs factorization theorem, while the dashed line is the prediction by the LP factorization.}
\end{figure}

At lower energies the situation is similar, but the shift is much larger. An example is shown in fig.\ref{fig:unpol-LE}, where we demonstrate the average of the first 5 bins in $q_T$ as a function of $Q$ (this average is consider in order to reduce statistical fluctuations). The curves are weighted by an exponential factor $Ne^{-Q}$ (where $N$ is tuned to the first experimental point) for a better visualization. In general, the theory underpredicts the data, although it reproduces the shape perfectly. This is a known feature of fixed target data, see refs.~\cite{Moos:2023yfa, Bacchetta:2022awv, Scimemi:2019cmh, Vladimirov:2019bfa, Gauld:2021pkr}. The prediction of the KPCs is considerably larger than the LP, from $5\%$ at $Q\sim 14$GeV up to $30\%$ at $Q\sim 5$GeV. Consequently, the KPCs do not completely solve the problem of the normalization of fixed-target data, but they do reduce the tension. With KPCs, the theory prediction is within the uncertainty window (systematic together with statistics) of the measurement.

\begin{figure}
\begin{center}
\includegraphics[width=0.32\textwidth]{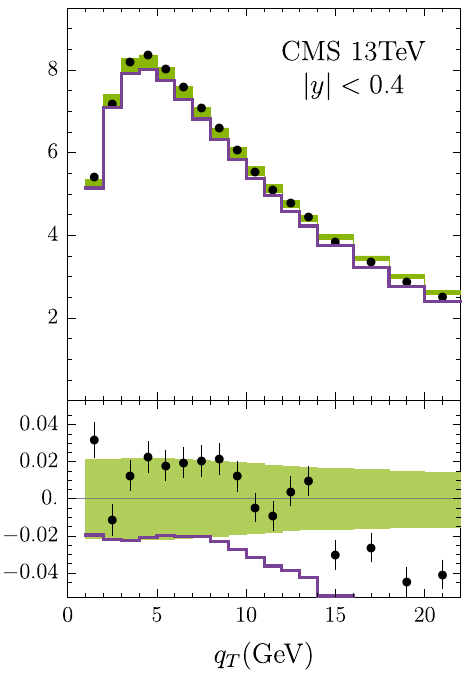}
\includegraphics[width=0.32\textwidth]{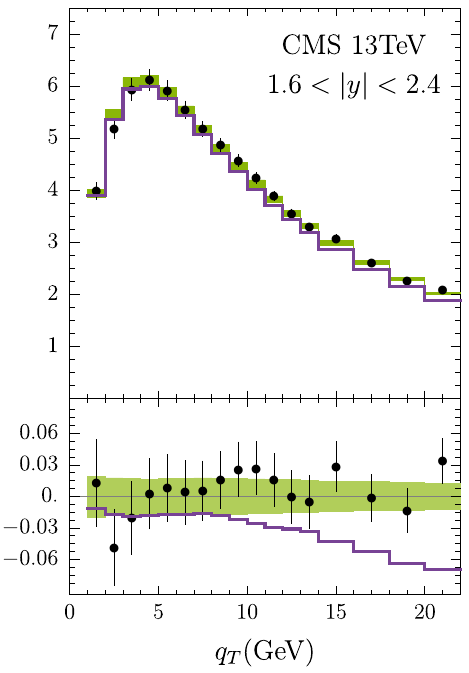}
\includegraphics[width=0.32\textwidth]{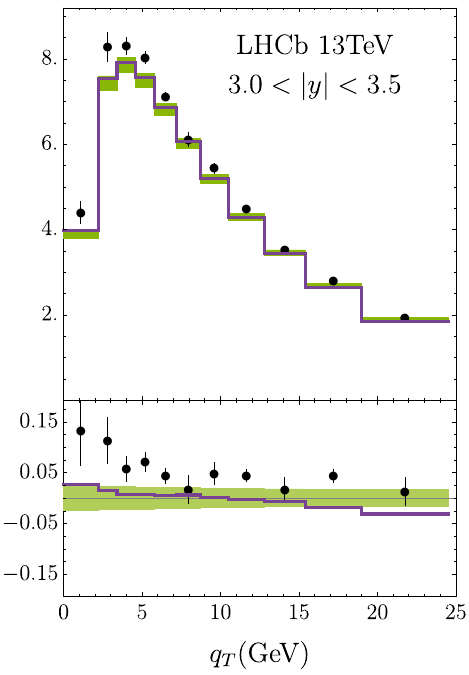}
\end{center}
\caption{\label{fig:unpol-HE-fiduc} Example of description of the fiducial cross-sections by only $\Sigma_U$ term, at different values of $y$. The band is the uncertainty of the theory prediction taken from ART23. The solid purple line is the LP prediction by ART23. The data comes from the measurements \cite{CMS:2019raw, LHCb:2021huf}. Note that the CMS and LHCb data have normalization uncertainties of $2.5\%$ and $2\%$, correspondingly.}
\end{figure}

The largest modifications take place for the fiducial cross-sections measured at the LHC. These measurements are the most accurate. Some of them reach a precision of $\sim 0.1\%$, so a tiny variation in the prediction leads to a large change in the $\chi^2$-value. For the fiducial cross-sections, the main modification in the theory prediction comes from the different expression for the fiducial-cut factor. The LP cut factor has a different behavior as a function of $q_T$, and thus the shape of the prediction is modified (see fig.\ref{fig:PLP_vs_PU}). Exactly this modification produces the largest rise of $\chi^2$. Apart from the fiducial factor, the prediction increases by $\sim 2\%$, which is a reasonable size for the contribution of the power corrections at LHC energies. Additionally, in fig.\ref{fig:unpol-HE-fiduc} we demonstrate the comparison with the measurements at $\sqrt{s}=13$TeV in different $y$-bins.

\subsection{Angular distribution $A_4$}

The angular distribution $A_4$ is the LP structure. It is unique in that it is proportional to the difference between the quark and anti-quark distributions, and hence sensitive to the difference between the quark and anti-quark transverse momenta. Presumably, this angular structure should be included in the standard extractions of $f_1$, especially in the light of the recent observation of flavor dependence of the transverse momentum \cite{Bury:2022czx, Moos:2023yfa, Bacchetta:2024qre}.

\begin{figure}
\begin{center}
\includegraphics[width=0.99\textwidth]{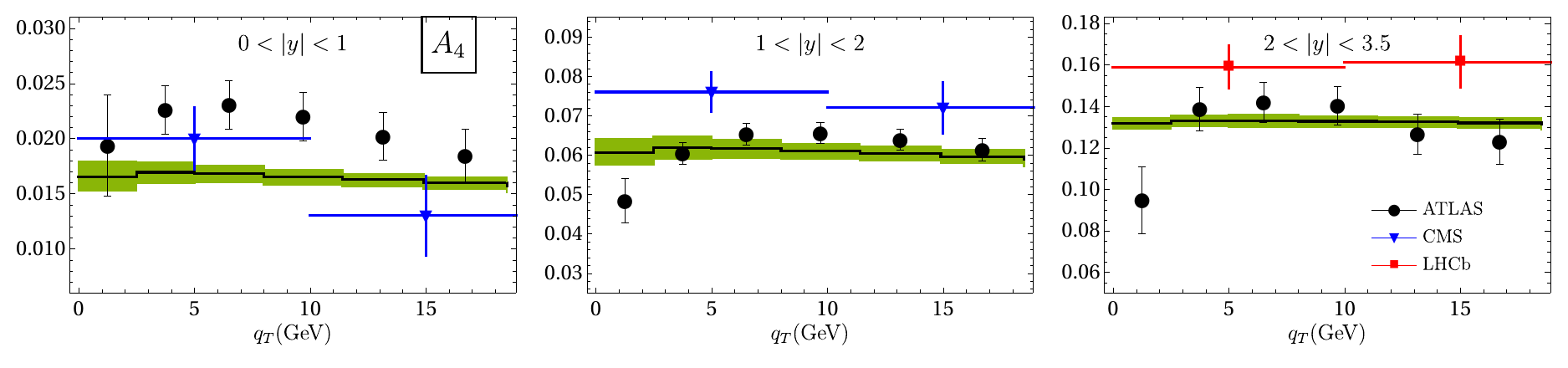}
\caption{\label{fig:A4:vsqT} Angular distribution $A_4$ as a function of $q_T$ vs. measurements of ATLAS \cite{ATLAS:2016rnf} (circles), CMS \cite{CMS:2015cyj} (triangles) and LHCb \cite{LHCb:2022tbc}.}
\end{center}
\end{figure}

The theory prediction based on the TMD factorization approach agrees very well with the measurements of $A_4$. The difference between the LP and TMD-with-KPCs factorization theorems is very small (of the order of 2-4\% for $A_4$). In fig.\ref{fig:A4:vsqT} we show the comparison of $A_4$ as a function of $q_T$ with ATLAS, CMS and LHCb measurements. The uncertainty band appears to be very small because the uncertainties of $\Sigma_U$ and $\Sigma_4$ are very correlated.

Note that the CMS and LHCb measurements are done in slightly different bins compared to ATLAS. For CMS the bins are $Q\in[81.,101]$GeV with $|y|\in[0,1]$ and $|y|\in[1,2.1]$. For LHCb the bin is $Q\in[75.,105.]$GeV and $y\in[2.,3.6]$. Also, the LHCb reports the value of $\Delta A_4$, which is the deviation of $A_4$ from its average value $\langle A_4\rangle_{\text{LHCb}}$. Meanwhile, LHCb does not report the value of $\langle A_4\rangle_{\text{LHCb}}$. Thus, we have computed it using our theory prediction, and we got $\langle A_4\rangle_{\text{LHCb}}=0.138\pm0.004$. We have added this theoretical uncertainty to the experimental one in the plot.

\begin{figure}
\begin{center}
\includegraphics[width=0.9\textwidth]{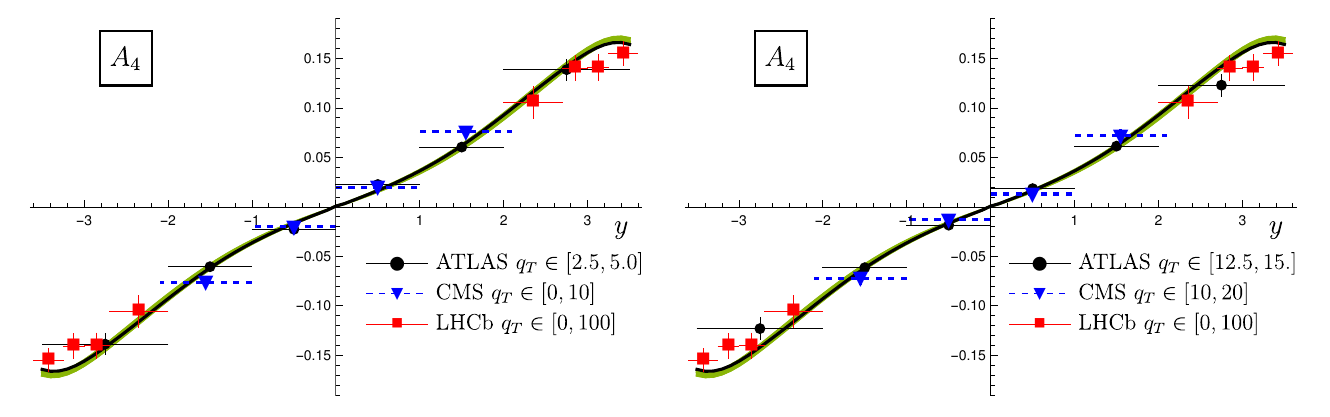}
\caption{\label{fig:A4:vsy} Angular distribution $A_4$ as a function of $y$ vs. measurements of ATLAS \cite{ATLAS:2016rnf} (circles), CMS \cite{CMS:2015cyj} (triangles) and LHCb \cite{LHCb:2022tbc}. The measurements are done in the $q_T$-bins indicated in the plots. The theory prediction is done for the $q_T$ bins of ATLAS.}
\end{center}
\end{figure}

The agreement between data and theory is even more transparent when $A_4$ is plotted as a function of $y$. In fig.\ref{fig:A4:vsy} we demonstrate such a comparison. Here we use that $A_4$ is almost constant with respect to $q_T$ (see fig.\ref{fig:A4:vsqT}) and therefore we compare the data points collected in different ranges of $q_T$. This comparison is not absolutely legitimate because all measurements are made on lightly different kinematics, as discussed above, but even so the agreement is spectacular. Note that the y-differential measurement of LHCb is done in a single $q_T$-bin, $q_T\in[0,100]$GeV.

The measurement of $A_4$ by CDF \cite{CDF:2011ksg} is done for the $p+\bar p$ system and is proportional to a different combination of TMD distributions. Computing the bins $q_T\in[0.,10.]$GeV and $q_T\in[10.,20]$GeV, we obtain $A_4$ as $0.121\pm0.006$ and $0.127\pm0.004$, respectively. These numbers are in agreement with the values $0.110\pm0.010$ and $0.101\pm0.017$ reported in ref.~\cite{CDF:2011ksg}, correspondingly.

\subsection{Angular distribution $A_2$}

\begin{figure}
\begin{center}
\includegraphics[width=0.99\textwidth]{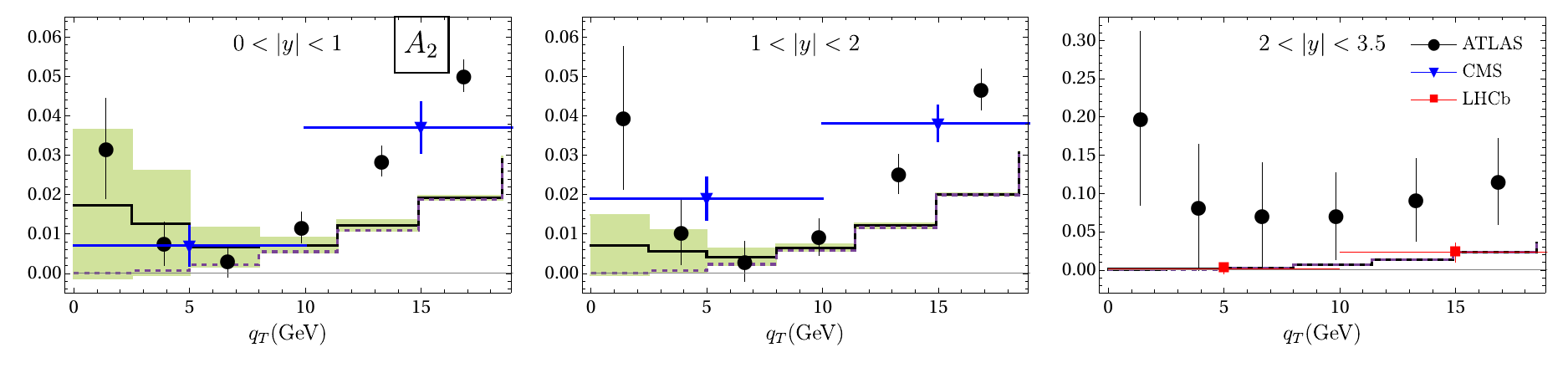}
\caption{\label{fig:A2:vsqT} Angular distribution $A_2$ as a function of $q_T$ vs. measurements of ATLAS \cite{ATLAS:2016rnf} (circles), CMS \cite{CMS:2015cyj} (triangles) and LHCb \cite{LHCb:2022tbc}. The dashed line shows the prediction without the Boer-Mulders contribution.}
\end{center}
\end{figure}

The Angular distribution $A_2$ contains the Boer-Mulders $\sim h_1^\perp h_1^\perp$ and the unpolarized $\sim f_1f_1$ contributions. They behave differently as functions of $q_T$. The Boer-Mulders part is LP, and thus it is dominant at $q_T\to0$. However, it drops rapidly as $q_T$ grows, since $h^\perp_1\sim 1/k_T^4$. The unpolarized contribution has opposite behavior -- it is suppressed as $q_T^2/Q^2$ at $q_T\to0$, and increases at larger $q_T$. The unpolarized contribution remains small due to the general power suppression $\sim M^2/Q^2$. Nonetheless, these two contributions are of the same general order, because the Boer-Mulders function is smaller than the unpolarized distribution by an order of magnitude.

The double-term structure is clearly visible in the ATLAS measurement, see fig.\ref{fig:A2:vsqT}. We associate the growth visible in the bins with $q_T<5$GeV with the Boer-Mulders term, while the growth at $q_T>7.5$GeV with the contribution of the unpolarized distributions. Note that at $q_T>10$GeV one expects an additional contribution from $q_T^2/Q^2$ corrections (Y-term). Moreover, we can quantify the general size of the Y-term correction using the ATLAS measurement as a baseline. It is $\sim 0.02$ at $q_T=14$GeV and $\sim 0.04$ at $q_T=18$GeV. This corresponds to $2\%$ and $4\%$ corrections for $\Sigma_U$, which agrees with general expectations. The same size discrepancy can be seen in $A_0$ and $A_1$.

As described in sec.~\ref{sec:h1}, we have fit the free parameters of the Boer-Mulders function $h_1^\perp$ (\ref{h1:model}) to the $A_2$ data by ATLAS at $q_T<10$GeV (12 points), resulting in $\chi^2/N_{pt}=1.16$. Notice that the unpolarized term contributes essentially to the $q_T>5$GeV region. The elimination of this term increases $\chi^2/N_{pt}$ to $1.9$.

It is important to emphasize that the double-Boer-Mulders contribution observed at ATLAS is positive. In this region of $Q$, the process is $Z$-boson dominated with the negative coupling constant $r_+^{ZZ}<0$ (see table \ref{tab:EW-constants}). Thus, in order to get a positive contribution, the $h_{1q}^\perp h_{1\bar q}^\perp$-term should be negative, which demands a negative relative sign between $h_{1q}^\perp$ and $h_{1\bar q}^\perp$. We also note that the $h_1^\perp=0$ case is not completely excluded by the ATLAS data and results in $\chi^2/N_{\text{pt}}=1.8$ (for 12 data points).

The large-rapidity measurement ($2<y<3.5$) is the most problematic. It is exclusively sensitive to the large-$x$ value of $h_1^\perp$. Our simplistic model produces a negligible $h_1^\perp$ for $x>0.3$, and, consequently, a negligible contribution to the large-rapidity bin. This is somewhat inconsistent with the ATLAS measurement, although in agreement with the one of the LHCb. Given the large uncertainty of the ATLAS measurement in this region, we cannot conclude a tension between theory and the data.

The measurements by other experiments, namely, CMS, LHCb and CDF, are done in the wide $q_T$-bins and are therefore insensitive to the double-term structure of factorized expressions. As it follows from the analysis of the ATLAS data, only the first bin $q_T\in[0,10]$GeV can be described within the TMD factorization. Already the second bin $q_T\in[10,20]$GeV has a significant contribution of power corrections. Interestingly, due to the $p+\bar p$-system, the Boer-Mulders contribution to the CDF measurement is negative, since $h_{1q}^\perp h_{1q}^\perp>0$. This term cancels almost identically the $f_1f_1$-term, resulting in a prediction of $-0.001\pm 0.004$ for the lowest $q_T$-bin (vs. measured $0.016\pm 0.026$ \cite{CDF:2011ksg}). The comparison with the CDF data is presented in table \ref{tab:CDF}.

\begin{table}[b]
\begin{center}
\begin{tabular}{|c||c|c||c|c|}
\hline
& \multicolumn{2}{c||}{CDF}  & \multicolumn{2}{c|}{Theory} 
\\
& $q_T\in[0,10]$ & $q_T\in[10,20]$ & $q_T\in[0,10]$ & $q_T\in[10,20]$
\\\hline
$A_0 (\times10^{})$  &  $0.17 \pm 0.16$ & $0.42\pm0.26$ & $0.05\pm 10^{-3}$ & $0.02\pm 10^{-3}$
\\\hline
$A_2 (\times10^{})$ &  $0.16 \pm 0.26$ & $-0.01\pm0.38$ & $0.01\pm 0.04$ & $0.15\pm 0.01$
\\\hline
$A_3 (\times10^{})$ &  $-0.04 \pm 0.12$ & $0.18\pm0.16$ & $-0.006\pm 0.036$ & $-0.05\pm 0.01$
\\\hline
$A_4 (\times10^{})$  &  $1.10 \pm 0.10$ & $1.01\pm0.17$ & $1.21 \pm 0.06$ & $1.27\pm0.04$
\\\hline
\end{tabular}
\caption{\label{tab:CDF} Comparison of the measurement by CDF \cite{CDF:2011ksg} with the prediction of the TMD factorization.}
\end{center}
\end{table}

\subsection{Angular distributions $A_0$, $A_1$, and $A_3$}

\begin{figure}
\begin{center}
\includegraphics[width=0.99\textwidth]{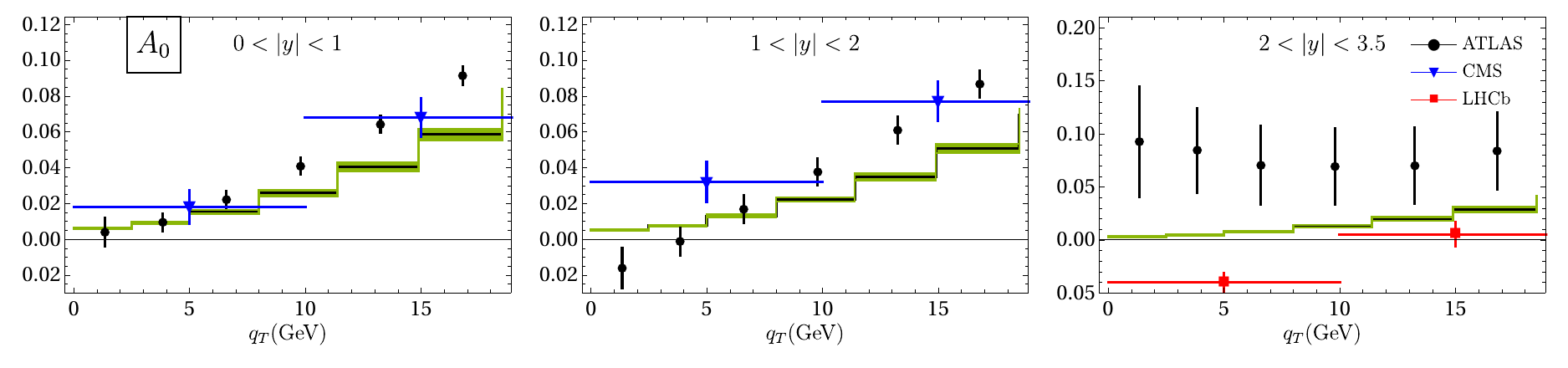}
\\
\includegraphics[width=0.99\textwidth]{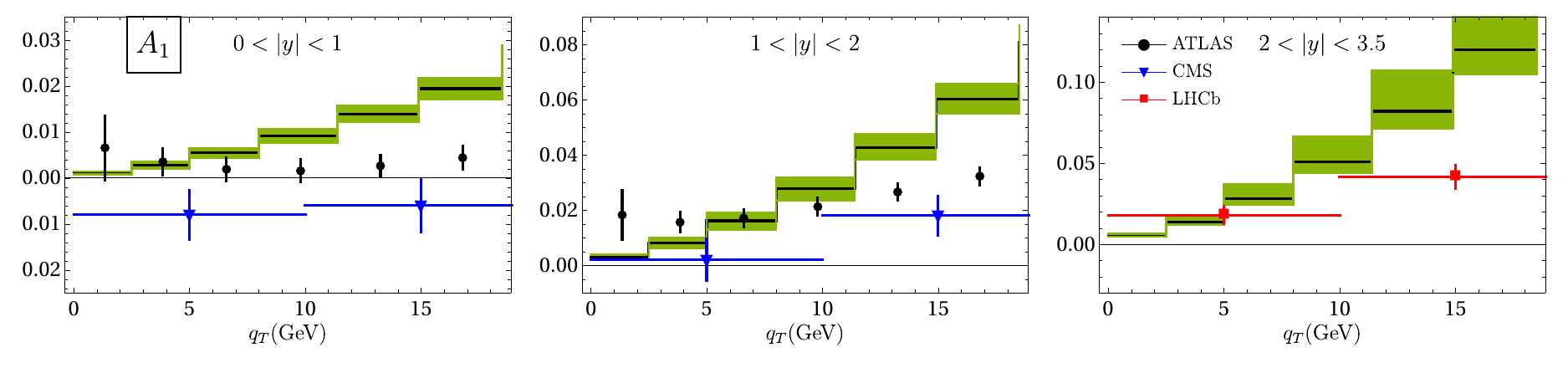}
\\
\includegraphics[width=0.99\textwidth]{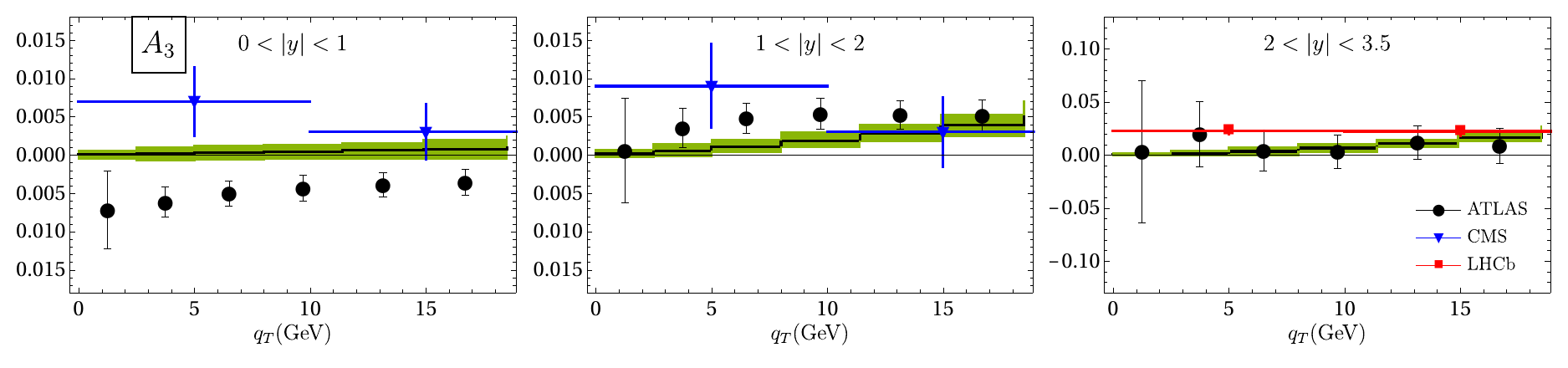}
\caption{\label{fig:A023} Angular distributions $A_0$, $A_1$ and $A_3$ as a function of $q_T$ vs. measurements of ATLAS \cite{ATLAS:2016rnf} (circles), CMS \cite{CMS:2015cyj} (triangles) and LHCb \cite{LHCb:2022tbc}.}
\end{center}
\end{figure}

The angular distributions $A_0$ and $A_1$ are dominated by the unpolarized term. The Boer-Mulders contributions to $A_0$ and $A_1$ are $\sim 10^{-3}$ and $\sim 10^{-6}$, respectively, compared to the $f_1f_1$-term. The comparison with the data is shown in fig.\ref{fig:A023}. One can see that at $q_T>10$GeV the deviation between data and prediction grows, which is evidence of the Y-term. It has the same general size as for the $A_2$ case, i.e., $\sim 0.02$ at $q_T=14$GeV and $\sim 0.04$ at $q_T=18$GeV. For $q_T<10$GeV the agreement between data and the theory is satisfactory.

The angular distribution $A_3$ is structurally similar to $A_4$ but has extra power-suppression. The bins with $y>1$ are nicely described by the theory. Meanwhile, the bin $|y|<1$ is measured by CMS and ATLAS and they present contradictory results. Note that in the case of $A_3$ we see no significant deviation from the data at $q_T\sim20$GeV. Possibly, this indicates a cancellation between the $q_T$-corrections to $\Sigma_U$ and $\Sigma_3$.

In the table \ref{tab:CDF} we present the comparison of the theory prediction with the measurement by CDF \cite{CDF:2011ksg}. This measurement is complementary to the LHC measurement because of the different flavor composition. We have found a perfect agreement in the lowest $q_T$-bin, and some disagreement in the second $q_T$-bin for $A_{0,1}$. This comparison confirms our previous observations.

\subsection{Angular distributions $\mu$, $\nu$, and $\lambda$}
\label{sec:nu}

\begin{figure}
\begin{center}
\includegraphics[width=0.36\textwidth]{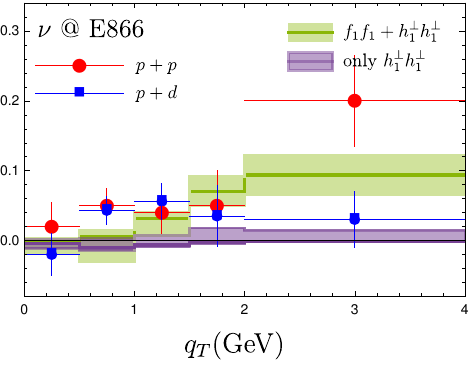}~~~~~~~
\includegraphics[width=0.365\textwidth]{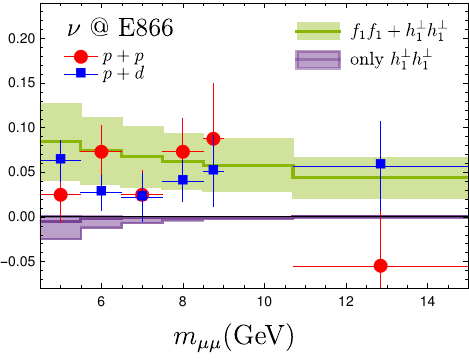}
\\
\includegraphics[width=0.36\textwidth]{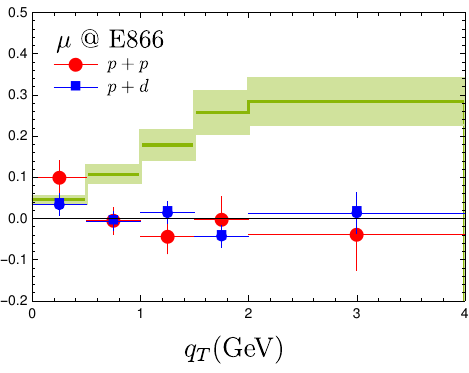}~~~~~~~
\includegraphics[width=0.36\textwidth]{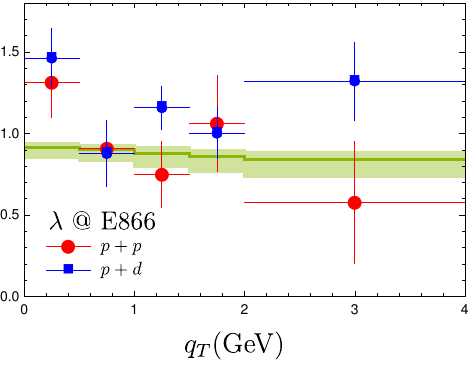}
\caption{\label{fig:nu} Comparison of the TMD factorization prediction with the measurement of the NuSea experiment \cite{NuSea:2008ndg}. The circle (square) markers represent the measurement in $p+p$ ($p+d$) collision. The theory prediction is computed for $p+p$ case. (Upper row) Angular distribution $\nu$ vs $q_T$ (left) and $Q$ (right). The theory prediction is shown for the complete expression, and for the $\sim h_1^\perp h_1^\perp$ contribution only. (Lower row) Angular distributions $\mu$ (left) and $\lambda$ (right).}
\end{center}
\end{figure}

The low-energy measurement of angular coefficients is presented by NuSea with the E866 experiment \cite{NuSea:2008ndg}. However, the description of these data faces significant problems due to the choice of binning. The measurement is performed in the range $Q\in[4.5,9.0]\cup [10.7,15.0]$GeV and $q_T\in[0,4]$GeV. Thus, it covers both TMD $(q_T\ll Q)$ and collinear $(q_T\sim Q)$ factorization regions. The data is grouped in several ways: $Q$-differential, $q_T$-differential and $x_{1,2}$-differential. The $q_T$-differential measurement is essentially dominated by the $Q\sim 5$GeV, because the cross-section grows rapidly at low-Q. Consequently, only low$-q_T$ bins (with $q_T<1-2$GeV) are within the TMD factorization region. In this case, one can also expect large target-mass and higher-twist corrections, since their typical size of $\sim M^2/Q^2 \sim 0.06$ is of the same order as the angular distributions. In the case of the $Q$-differential measurement, the high-$Q$ bins (with $Q>8-10$GeV) are in the TMD factorization region. The $x_{1,2}$-differential measurements represent a mix of all factorization regions, and could not be reliably studied within a single-factorization approach. Thus, we do not expect a good agreement for any of these measurements, except for $Q>8-10$GeV.

The angular structure decomposition used by NuSea is different. It reads
\begin{eqnarray}
\frac{d\sigma}{d\Omega}\propto 1+\lambda \cos^2\theta+\mu\sin 2\theta\cos\phi+\frac{\nu}{2}\sin^2\theta\cos 2\phi.
\end{eqnarray}
Comparing with (\ref{def:angular}) we find the relation between the distributions $\mu$, $\nu$ and $\lambda$ and the structure functions $\Sigma_i$,
\begin{eqnarray}
\lambda=\frac{2\Sigma_U-3\Sigma_0}{2\Sigma_U+\Sigma_0},
\qquad
\mu=\frac{2\Sigma_1}{2\Sigma_U+\Sigma_0},
\qquad
\nu=\frac{2\Sigma_2}{2\Sigma_U+\Sigma_0}.
\end{eqnarray}
The comparison of these functions with the NuSea measurement is shown in fig.\ref{fig:nu}. The measurement is done for $p+p$ and $p+d$ collisions. In our study, we have neglected the differences coming from the flavor decomposition and only compare the $p+p$ case.

We observe that the angular distributions $\nu$ and $\lambda$ are in general agreement with the measurement. However, the distribution $\mu$ deviates significantly. The theory agrees with the experiment at $q_T<0.5$GeV and then overgrows the data. We associate this deviation with large power corrections of the $Y$-term which become significant at $q_T>1$GeV. Moreover, the analogous distribution $A_1$, shows a similar behavior (the prediction overgrows the measurement).

We emphasize that the angular distribution $\nu$ in this kinematic regime is dominated by the $f_1f_1$-term, despite it is formally of sub-leading power. Eliminating this term leads to a small and negative $\nu$ (see the purple band in fig.\ref{fig:nu} (upper row)), which does not describe the data. The negativity of the $h_1^\perp h_1^\perp$ term at low-$Q$ follows from its positivity at $Q\sim M_Z$, because $r^{\gamma\gamma}_{+q}>0$. So, the $h_1^\perp h_1^\perp$-term changes sign during the evolution from low to high $Q$ at $Q\sim 70$GeV. This effect was also discussed in ref.~\cite{Lu:2011mz}.

There are exact inequalities that follow from the positive-definiteness of the hadron tensor for the electro-magnetic current \cite{Lam:1978pu}. In terms of the structure functions $\Sigma_i$ they read
\begin{eqnarray}
\Sigma_1\geqslant 0,\qquad 
2 \Sigma_U\geqslant|\Sigma_2|+\Sigma_0,\qquad 
\Sigma_U\Sigma_0\geqslant (\Sigma_0+\Sigma_2)\Sigma_0 +\Sigma_1^2.
\end{eqnarray}
Note that the first relation sets a model-independent constraint on the Boer-Mulders function which has an integral form, and could not be revealed explicitly. However, using the large-$Q$ form $\Sigma_0$, we can present it in its approximate form
\begin{eqnarray}
f_1(x_1,\vec k_1)f_1(x_2,\vec k_2)\gtrsim-\frac{(\vec k_1-\vec k_2)^2}{4M^2}h^\perp_1(x_1,\vec k_1)h^\perp_1(x_2,\vec k_2).
\end{eqnarray}
This is naturally satisfied, since $f_1$ is generally bigger than $h_1^\perp$, and decays as $k_T^{-2}$ vs. $k_T^{-4}$ in the case of the Boer-Mulders function. Note that the measurement of $\lambda$ at E866 violates this positivity constraint.

\subsection{Lam-Tung relation}

\begin{figure}
\begin{center}
\includegraphics[width=0.4\textwidth]{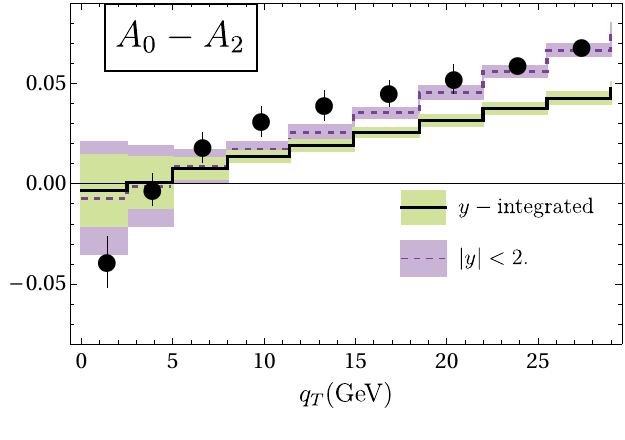}
~~~
\includegraphics[width=0.36\textwidth]{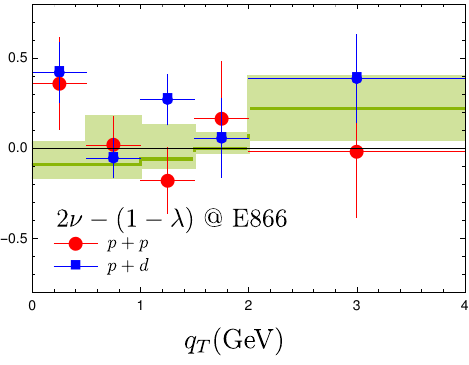}
\caption{\label{fig:LT} Lam-Tung combination of angular distributions measured at ATLAS \cite{ATLAS:2016rnf} (left) and NuSea \cite{NuSea:2008ndg}(right) vs $q_T$.}
\end{center}
\end{figure}

Another combination that attracts interest is the Lam-Tung relation \cite{Lam:1978pu, Lam:1978zr}. In the present notation the Lam-Tung combination reads
\begin{eqnarray}
\Sigma_{\text{LT}}=\Sigma_2-\Sigma_0.
\end{eqnarray}
In the large-$q_T$ regime, the collinear factorization predicts that $\Sigma_{\text{LT}}\sim \mathcal{O}(\alpha_s^2)$, because the leading perturbative contributions to $\Sigma_{0,2}$ cancel. In the TMD factorization theorem, the Lam-Tung relation does not hold already at LP and LO. However, by combining (\ref{def:S0}) and (\ref{def:S2}), many terms cancel producing a simpler expression
\begin{eqnarray}\label{def:LT}
&&\Sigma_{\text{LT}}=
\frac{4\pi \alpha_{\text{em}}^2}{3N_cs}\sum_{q,G,G'} Q^4\Delta_G^*\Delta_{G'}\Big\{
z_{+\ell}^{GG'}z_{+q}^{GG'}\mathcal{C}[2\frac{((\vec  t\vec k_1)-(\vec t\vec k_2))^2-(\vec k_1-\vec k_2)^2}{Q^2},f_1f_1]
\\ && \nn
+z_{+\ell}^{GG'}r_{+q}^{GG'}\mathcal{C}[
\frac{\vec k_1^2+\vec k_2^2-((\vec  t\vec k_1)-(\vec t\vec k_2))^2}{M^2} 
+\frac{\vec k_1^2+\vec k_2^2}{M^2}\frac{((\vec  t\vec k_1)-(\vec t\vec k_2))^2-(\vec k_1-\vec k_2)^2}{Q^2}
,h^\perp_1h^\perp_1]\Big\}.
\end{eqnarray}
It is worth mentioning that the most power-suppressed $\sim Q^{-4}$ parts cancel entirely, and the $\sim Q^{-2}$ parts have the same general prefactor for the $f_1f_1$ and $h_1^\perp h_1^\perp$ contributions. 

In fig.\ref{fig:LT} we demonstrate the comparison of the measurement of the Lam-Tung relation made at ATLAS and NuSea. In the latter case, the following combination is presented
\begin{eqnarray}
2\nu-(1-\lambda)=\frac{4\Sigma_{\text{LT}}}{2\Sigma_U+\Sigma_0}.
\end{eqnarray}
In the case of ATLAS, the TMD factorization prediction is almost two times smaller than the data. This is clearly an effect of the imperfection of the model at large-$x$. Indeed, the data indicates a rather flat behavior in $y$ \citep{ATLAS:2016rnf}, while the theory drops to zero at large-$y$. If we eliminate this problematic region and restrict $y\in[0,2]$ we have a good agreement with the data. 

The comparison with the CMS and LHCb measurements is given in the table \ref{tab:LT}. These data are differential in $y$ and provide a further justification for our observation. The agreement with the CMS data, that is collected for $y<2$ is very good. Meanwhile, there is a significant disagreement with the LHCb data. Thus, we conclude that the current TMD models incorrectly describe the TMD distributions at large-x. Note that this problem has also been pointed out in other studies, see, e.g., \cite{delRio:2024vvq, Vladimirov:2019bfa}.

The Lam-Tung relation has the Y-term suppressed by an extra factor $\alpha_s$. Therefore, the power corrections associated with the Y-term should be smaller. Indeed, the Lam-Tung relation (in the region of $y<2$) is in a very good agreement with the prediction of TMD factorization up to $q_T\sim 30$GeV in contrast to $A_0$ and $A_2$ independently. 

The good description of the Lam-Tung relation is only possible due to the inclusion of KPCs into TMD factorization. The LP TMD factorization theorem predicts $\Sigma_{LP}\sim \vec k^2/M^2 h_1^\perp h_1^\perp$, which is leads to $A_0-A_2<10^{-3}$ for $q_T>10$GeV. Therefore, the LP factorization essentially disagrees with the data, while the TMD-with-KPCs expression is in a good agreement.

\begin{table}[b]
\begin{center}
\begin{tabular}{|c||c|c|c|c|c|c||}
\hline
$|y|$-range & $[0,1]$ & $[0,1]$ & $[0,1]$ & $[1,2.1]$ & $[1,2.1]$ & $[1,2.1]$
\\
$q_T$-range (GeV) & $[0,10]$ & $[10,20]$ & $[20,35]$ & $[0,10]$ & $[10,20]$ & $[20,35]$
\\\hline
CMS \cite{CMS:2015cyj} $(\times 10^{2})$ & $1.1\pm 1.0$ & $3.2\pm 1.3$ & $4.3\pm 1.7$ & $1.3\pm 1.3$ & $3.9\pm 1.4$ & $5.1\pm 2.0$ 
\\\hline
Theory $(\times 10^{2})$ & $0.3\pm 1.8$ & $3.3\pm 0.6$  & $7.3\pm 0.3$ & $0.6\pm 0.8$ & $2.5\pm0.5$ & $ 5.2\pm 0.2$
\\\hline
\end{tabular}
\begin{tabular}{|c||c|c|c||}
\hline
$q_T$-range (GeV) & $[0,10]$ & $[10,20]$ & $[20,35]$
\\\hline
LHCb \cite{LHCb:2022tbc} $(\times 10^{2})$ & $-4.2\pm1.1$ & $ -4.5\pm1.4$ & $2.9\pm 0.2$
\\\hline
Theory $(\times 10^{2})$ &  $0.3\pm0.2$ & $ -0.08\pm0.41$ & $-0.3\pm 0.2$
\\\hline
\end{tabular}
\caption{\label{tab:LT} Comparison of the measurements of the Lam-Tung relation by CMS \cite{CMS:2015cyj} (upper table) and LHCb \cite{LHCb:2022tbc} (lower table) with the prediction of the TMD factorization.}
\end{center}
\end{table}

\section{Conclusion}

In this work we have studied the angular coefficients of the Drell-Yan lepton pair produced by a neutral electro-weak boson. The analysis is done in the framework of the transverse momentum dependent (TMD) factorization theorem, with the inclusion of kinematic power corrections (KPCs), as it was suggested in ref.~\cite{Vladimirov:2023aot}. This implementation of the factorization theorem includes all power-suppressed terms containing TMD distributions of twist-two at small values of $q_T$. The inclusion of KPCs restores the gauge- and frame-invariance of the hadron tensor, which are broken at the leading power (LP) approximation, and is crucial in many other aspects. In particular, it renders non-zero angular structure functions that vanish at LP (except for the angular coefficient $A_7$, which is thus a pure higher-twist term). One of the successes of the TMD-with-KPCs factorization approach is a good description of the Lam-Tung relation, for which the LP approximation essentially disagrees with the data. 

The TMD factorization theorem with included KPCs preserves the perturbative and non-perturbative content of the factorization expression at LP. This means that it uses the same parton distributions and the same coefficient functions. In practice, the main modification is the expression for the integral convolution of the distributions (discussed in detail in appendix \ref{app:convolution}). Therefore, we are able to utilize all known perturbative orders and made the analysis at N$^4$LL (with NNLO PDF input for the unpolarized distribution) alike in ref.~\cite{Moos:2023yfa}. 

We have implemented the KPCs in \texttt{artemide}. This required a significant update of the code. The main modifications were made in the Fourier procedure, in the computation of the cut-factors and in the convolution integral. They are discussed in detail in the appendix \ref{sec:implementation}. The new version (version 3) of \texttt{artemide} is available at \cite{artemide}.

All angular coefficients are described in terms of the Boer-Mulders and the unpolarized TMD distributions. The unpolarized TMD distribution is taken from ART23 \cite{Moos:2023yfa}, with minimal modifications (described in sec.~\ref{sec:NP-setup}). The Boer-Mulders function is determined using the data for $A_2$ angular distribution measured by ATLAS \cite{ATLAS:2016rnf}. We argue that the peculiar shape of $A_2$ visible at low-$q_T$ (see fig.\ref{fig:A2:vsqT}) is an evidence of the Boer-Mulders contribution. There are few data-points in this region, so we can only identify a general size of the Boer-Mulders function. It is interesting that the data points toward the opposite-signs of the quark and anti-quark distributions. In the collinear limit, the Boer-Mulders function turns into a twist-three $E(-x,0,x)$-distribution. To our best knowledge, this is the first observation of high-twist effects at the LHC.

In general, we have found a satisfactory agreement between the theoretical prediction and the data. The main part of the data comes from measurements at ATLAS \cite{ATLAS:2016rnf}, but also from CMS \cite{CMS:2015cyj}, LHCb \cite{LHCb:2022tbc} and CDF \cite{CDF:2012brb}. The low-energy measurement made by the NuSea \cite{NuSea:2008ndg} experiment is somewhat problematic, due to the inappropriate kinematic region for the application of the TMD factorization theorem (see discussion in sec.~\ref{sec:nu}). We emphasize that in all cases (low- and high-energy measurements) it is important to consider both unpolarized and Boer-Mulders contributions, despite one of them may be power suppressed. For example, in the angular distribution $A_2$ (or $\nu$), both terms contribute at the same numerical order, because the Boer-Mulders function is smaller than the unpolarized one, while the unpolarized term is accompanied by the power-suppressed factor. This results into an involved dependence on $Q$, since these terms evolve differently and have different couplings to $\gamma$ and $Z$-bosons.

The present analysis also points out the problem of describing large-$x$ data in the TMD factorization theorem. The potential presence of these problems was discussed in ref.~\cite{Vladimirov:2019bfa, Cao:2021aci, Cerutti:2022lmb, Bury:2022czx}. However, the analysis of the angular coefficients shows it more explicitly. Generally, all measurements at $|y|>2$ are underestimated by $5-10\%$.

The TMD-with-KPCs factorization theorem has the same kinematic region of applicability as the LP TMD factorization theorem. I.e. $Q\to\infty$ and $q_T\ll Q$. For larger-$q_T$, other types of power corrections start to contribute (mainly $q_T/Q$-corrections in the nomenclature of \cite{Vladimirov:2023aot}). The appearance of power corrections is clearly visible in the comparison of the angular coefficients with the data, which allows us to conclude that power corrections are $\sim 2\%$ are $q_T\sim14$GeV and $\sim 4\%$ at $q_T\sim 18$GeV. This observation is in agreement with other studies, such as \cite{Scimemi:2019cmh, Bacchetta:2022awv}. For the Lam-Tung relation, where the large-$q_T$ tale is known to be extra suppressed by $\alpha_s$, the discrepancy is correspondingly smaller. This allows us to provide a prediction for the Lam-Tung up to higher values of $q_T$.

This work is the first practical application of the TMD-with-KPCs factorization theorem. It demonstrated a very good description of the data and sensitivity to a previously unreachable physical structures. It requires a fine-tuning of the TMD distributions, which were obtained in the LP approximation, but it has a good predictive power even without it.

\acknowledgments

We are very thankful to P.Zurita, C.van Hulsen, I.Scimemi, and A.Prokudin for multiple discussions and helpful remarks. This project is supported by the Atracci\'on de Talento Investigador program of the Comunidad de Madrid (Spain) No. 2020-T1/TIC-20204 and Europa Excelencia EUR2023-143460, MCIN/AEI/10.13039/501100011033/, from Spanish Ministerio de Ciencias y Innovaci\'on.

\appendix

\section{Coupling constants and propagators for neutral bosons}
\label{app:couplings}

The computation of Z/$\gamma$-boson production involves the current
\begin{eqnarray}
\gamma_{G}^\mu =g_R^{G}\gamma^\mu(1+\gamma^5)+g_L^{G}\gamma^\mu(1-\gamma^5),
\end{eqnarray}
where $g_R^G$ and $g_L^G$ represent the right and left electroweak coupling constants, respectively, and $G$ denotes the type of gauge boson, which, in our case, is either $\gamma$ or $Z$. Alternatively, one can also write the current in terms of the vector $v_f$ and axial $a_f$ couplings
\begin{eqnarray}
\gamma_G^\mu=\frac{1}{2s_Wc_W}\gamma^\mu (v_f-a_f\gamma^5),
\end{eqnarray}
which are related to $g_R^G$ and $g_L^G$ through the following equations
\begin{eqnarray}
v_f=2s_Wc_W(g_{R}^{G}+g_{L}^{G}),\qquad
a_f=2s_Wc_W(g_{L}^{G}-g_{R}^{G}).
\end{eqnarray}

As we noted in sec.~\ref{sec:theory}, it is useful to introduce the subsequent combinations of these coupling constants when formulating the decomposition of the lepton tensor (\ref{L:def}), and, consequently, when computing the angular structure functions $\Sigma_n$ (\ref{def:SU}-\ref{def:S7})
\begin{eqnarray}\label{def:zP}
z_{+}^{GG'}&=&2(g_{R}^{G}g_{R}^{G'}+g_{L}^{G}g_{L}^{G'})=\frac{v^G v^{G'}+a^G a^{G'}}{4 s_W^2c_W^2},
\\ \label{def:rP}
r_{+}^{GG'}&=&2(g_{R}^{G}g_{L}^{G'}+g_{L}^{G}g_{R}^{G'})=\frac{v^G v^{G'}-a^G a^{G'}}{4 s_W^2c_W^2},
\\ \label{def:zM}
z_{-}^{GG'}&=&2(g_{R}^{G}g_{R}^{G'}-g_{L}^{G}g_{L}^{G'})=-\frac{v^G a^{G'}+a^G v^{G'}}{4 s_W^2c_W^2},
\\ \label{def:rM}
r_{-}^{GG'}&=&2(g_{R}^{G}g_{L}^{G'}-g_{L}^{G}g_{R}^{G'})=\frac{v^G a^{G'}-a^G v^{G'}}{4 s_W^2c_W^2}.
\end{eqnarray}
The specific expressions for $g_R^G$ and $g_L^G$, particularized for the bosons under consideration, are
\begin{eqnarray}\label{explicit_g}
g_R^Z=\frac{-e_fs_W^2}{2s_Wc_W},\qquad g_L^Z=\frac{T_3-e_fs_W^2}{2s_Wc_W},\qquad g_\gamma^R=g_\gamma^L=\frac{e_f}{2},
\end{eqnarray}
where $e_f$ is the electric charge of a fermion $f$ (in units of $e$), $T_3$ is the third projection of its weak isospin, and $s_W$ and $c_W$ are the sine and cosine of the Weinberg angle, $\theta_W$, i.e., $s_W = \sin{(\theta_W)}$ and $c_W = \cos{(\theta_W)}$. All possible coupling constants combinations for the neutral bosons we are interested in can be obtained by substituting (\ref{explicit_g}) into (\ref{def:zP})-(\ref{def:rM}) 
\begin{eqnarray}
&&z_{+f}^{\gamma\gamma}=r_{+f}^{\gamma\gamma}=e_f^2,
\\\nn
&&z_{-f}^{\gamma\gamma}=r_{-f}^{\gamma\gamma}=0,
\\
&&z_{+f}^{\gamma Z}=z_{+f}^{Z\gamma}=r_{+f}^{\gamma Z}=r_{+f}^{Z\gamma}=\frac{e_f(T_3-2e_fs_W^2)}{2s_Wc_W} = \frac{|e_f|-4e_f^2s_W^2}{4s_wc_W},
\\
&&z_{+f}^{ZZ}=\frac{T_3^2-2e_fT_3s_W^2+2e_f^2s_W^4}{2s^2_Wc^2_W}=\frac{(1-2|e_f|s_W^2)^2+4e_f^2s_W^4}{8s_W^2c_W^2},
\\
&&r_{+f}^{ZZ}=\frac{e_f(e_fs_W^2-T_3)}{c_W^2}=\frac{2e^2_fs^2_W-|e_f|}{2c_W^2},
\\
&&z_{-f}^{\gamma Z}=z_{-f}^{Z\gamma}=-r_{-f}^{\gamma Z}=r_{-f}^{Z\gamma}=-\frac{T_3e_f}{2s_Wc_W}=-\frac{|e_f|}{4s_Wc_W},
\\
&&z_{-f}^{ZZ}=\frac{T_3(2e_fs_W^2-T_3)}{2s_W^2c_W^2}=\frac{4|e_f|s_W^2-1}{8s_W^2c_W^2},
\\
&& r_{-f}^{ZZ}=0.
\end{eqnarray}
The numeric values of these constants are given in table \ref{tab:EW-constants}.

Finally, taking into account all the previously outlined possibilities and considering that the propagator of a neutral gauge boson is given by
\begin{eqnarray}
\Delta_G(Q) = \frac{\delta_{G\gamma}}{Q^2+i0} + \frac{\delta_{GZ}}{Q^2-M_Z^2 + i\Gamma_ZM_Z},
\end{eqnarray}
we can explicitly write the expressions for the combinations that appear within the angular structure functions $\Sigma_n$ (\ref{def:SU}-\ref{def:S7}), i.e.,
\begin{eqnarray}
&&\sum_{GG'}Q^4z_{+\ell}^{GG'}z_{+f}^{GG'}\Delta^{GG'}(Q^2)=
\\\nn &&
\qquad
z_{+\ell}^{\gamma\gamma}z_{+f}^{\gamma\gamma}
+
z_{+\ell}^{\gamma Z}z_{+f}^{\gamma Z}\frac{2Q^2(Q^2-M_Z^2)}{(Q^2-M_Z^2)^2+\Gamma_Z^2M_Z^2}
+
z_{+\ell}^{Z Z}z_{+f}^{Z Z}\frac{Q^4}{(Q^2-M_Z^2)^2+\Gamma_Z^2M_Z^2},
\\
&&\sum_{GG'}Q^4z_{+\ell}^{GG'}r_{+f}^{GG'}\Delta^{GG'}(Q^2)=
\\\nn &&
\qquad
z_{+\ell}^{\gamma\gamma}z_{+f}^{\gamma\gamma}
+
z_{+\ell}^{\gamma Z}z_{+f}^{\gamma Z}\frac{2Q^2(Q^2-M_Z^2)}{(Q^2-M_Z^2)^2+\Gamma_Z^2M_Z^2}
+
z_{+\ell}^{Z Z}r_{+f}^{Z Z}\frac{Q^4}{(Q^2-M_Z^2)^2+\Gamma_Z^2M_Z^2},
\\
&&\sum_{GG'}Q^4z_{-\ell}^{GG'}z_{-f}^{GG'}\Delta^{GG'}(Q^2)=
\\\nn &&\qquad
z_{-\ell}^{\gamma Z}z_{-f}^{\gamma Z}\frac{2Q^2(Q^2-M_Z^2)}{(Q^2-M_Z^2)^2+\Gamma_Z^2M_Z^2}
+
z_{-\ell}^{Z Z}z_{-f}^{Z Z}\frac{Q^4}{(Q^2-M_Z^2)^2+\Gamma_Z^2M_Z^2},
\\
&&\sum_{GG'}Q^4 iz_{+\ell}^{GG'}r_{-f}^{GG'}\Delta^{GG'}(Q^2)=
z_{+\ell}^{\gamma Z}r_{-f}^{\gamma Z}\frac{2Q^2 M_Z\Gamma_Z}{(Q^2-M_Z^2)^2+\Gamma_Z^2M_Z^2},
\end{eqnarray}
where we have denoted $\Delta^{GG'}(Q^2) = \Delta^*_G(Q)\Delta_{G'}(Q)$.

\begin{table}[t]
\begin{center}
\begin{tabular}{|c||c|c|c|c|c|c|c|c|c|}\hline
 & $z_{+f}^{\gamma\gamma}$ & $z_{-f}^{\gamma\gamma}$ & $z_{+f}^{\gamma Z}, z_{+f}^{Z \gamma}$ & $z_{-f}^{\gamma Z}, z_{-f}^{Z\gamma}$ & $z_{+f}^{ZZ}$ & & $z_{-f}^{ZZ}$ & \\
$f$ & $r_{+f}^{\gamma\gamma}$ & $r_{-f}^{\gamma\gamma}$ & $r_{+f}^{\gamma Z}, r_{+f}^{Z \gamma}$ & $-r_{-f}^{\gamma Z}, r_{-f}^{Z\gamma}$ & & $r_{+f}^{ZZ}$ & & $r_{-f}^{ZZ}$
\\\hline
$\ell$ & $1$ & 0 & $0.0445$ & $-0.5930$ & $0.3536$ & $-0.3496$ & $-0.0528$ &0
\\\hline
$u, c$ & $\frac{4}{9}$ & 0 & $0.1516$ & $-0.3953$ & $0.4033$ & $-0.2999$ & $-0.2696$ &0
\\\hline
$d, s, b$ & $\frac{1}{9}$ & 0 & $0.1367$ & $-0.1977$ & $0.5198$ & $-0.1834$ & $-0.4864$ &0
\\\hline
\end{tabular}
\caption{\label{tab:EW-constants} Values of the various coupling constants combinations for $s_W^2=0.23122$ \cite{ParticleDataGroup:2022pth}. }  
\end{center}
\end{table}

\section{Details on the implementation in \texttt{artemide}}
\label{sec:implementation}

\subsection{Cut-factors}
\label{app:cut-factors}

The lepton tensor with fiducial cuts is defined in (\ref{L-cut:definition}) as
\begin{eqnarray}\label{L-cut:app}
\widehat{L}^{\mu\nu}_{GG'}(\text{cuts})=\int \frac{d^3l}{2E}\frac{d^3l'}{2E'}L_{GG'}^{\mu\nu}\Theta(\text{cuts})\delta^{(4)}(q-l-l'),
\end{eqnarray}
where $\Theta$ is the multidimensional step function that encodes the condition for the leptons momenta to belong to the fiducial region, and $L_{GG'}^{\mu\nu}$ is the lepton tensor in the Born approximation (\ref{L_0})
\begin{eqnarray}\label{L_0:app}
L^{\mu\nu}_{GG'}&=&4\[z_{+\ell}^{GG'}(l^\mu l'^\nu+l'^\mu l^\nu-g^{\mu\nu}(ll'))
-iz_{-\ell}^{GG'} \epsilon^{\mu\nu\alpha\beta}l_\alpha l_\beta' \].
\end{eqnarray}
The decomposition into a set of independent tensors is given in (\ref{L-cut:decomposition}) and reads
\begin{eqnarray}\label{L:decomposition:app}
\widehat{L}^{\mu\nu}_{GG'}(\text{cuts})=\frac{-2\pi Q^2}{3}\(
z^{GG'}_{+\ell}\sum_{n=U,0,1,2,5,6}\mathcal{P}_n(\text{cuts})\mathfrak{L}_n^{\mu\nu}
+
z^{GG'}_{-\ell}\sum_{n=3,4,7}\mathcal{P}_n(\text{cuts})\mathfrak{L}_n^{\mu\nu}\),
\end{eqnarray}
where the tensors $\mathfrak{L}_n^{\mu\nu}$ are defined in eqns.(\ref{L:U} -- \ref{L:7}) and $\mathcal{P}_n$ denote the cut-factors we are interested in. In order to derive them, we contract tensors $\mathbf{P}^{\mu\nu}_n$ such that
\begin{eqnarray}
\mathbf{P}^{\mu\nu}_n\mathfrak{L}_{m,\mu\nu}=\delta_{nm}.
\end{eqnarray}
The expressions for these tensors are
\begin{align}\nn
&\mathbf{P}^{\mu\nu}_{U}=\frac{1}{2}\mathfrak{L}_U^{\mu\nu}-\frac{1}{4}\mathfrak{L}_0^{\mu\nu},
&&\mathbf{P}^{\mu\nu}_{0}=-\frac{1}{4}\mathfrak{L}_U^{\mu\nu}+\frac{3}{8}\mathfrak{L}_0^{\mu\nu},
&&\mathbf{P}^{\mu\nu}_{1}=\frac{1}{2}\mathfrak{L}_1^{\mu\nu},
\\
&\mathbf{P}^{\mu\nu}_{2}=\frac{1}{8}\mathfrak{L}_2^{\mu\nu},
&&\mathbf{P}^{\mu\nu}_{3}=\frac{-1}{8}\mathfrak{L}_3^{\mu\nu},
&&\mathbf{P}^{\mu\nu}_{4}=\frac{-1}{8}\mathfrak{L}_4^{\mu\nu},
\\\nn
&\mathbf{P}^{\mu\nu}_{5}=\frac{1}{2}\mathfrak{L}_5^{\mu\nu},
&&\mathbf{P}^{\mu\nu}_{6}=\frac{1}{2}\mathfrak{L}_6^{\mu\nu},
&&\mathbf{P}^{\mu\nu}_{7}=\frac{-1}{8}\mathfrak{L}_7^{\mu\nu}.
\end{align}
Comparing (\ref{L-cut:app}) and (\ref{L:decomposition:app}) and one finds
\begin{eqnarray}\label{P-general:app}
\mathcal{P}_n(\text{cuts})=\frac{-3}{2\pi Q^2}\frac{1}{z^{GG'}_{\pm \ell}}\int \frac{d^3l}{2E}\frac{d^3l'}{2E'}\Theta(\text{cuts})\delta^{(4)}(q-l-l')
L_{GG'}^{\mu\nu}\mathbf{P}_{n,\mu\nu},
\end{eqnarray}
where $z^{GG'}_{\pm \ell}$ is $z^{GG'}_{+\ell}$ for $n=U,0,1,2,5,6$ and $z^{GG'}_{-\ell}$ for $n=3,4,7$.

A typical experimental measurement imposes the following constraints
\begin{eqnarray}\label{L:cut-region1:app}
|\vec l_T|>p_{1T},\qquad |\vec l'_T|>p_{2T},\qquad \eta_{\text{min}}<\eta,\eta'<\eta_{\text{max}},
\end{eqnarray}
where $\eta$ and $\eta'$ are the leptons rapidities. To incorporate these restrictions into (\ref{P-general:app}) we use the subsequent parametrization for the vectors
\begin{eqnarray}\nn
n&=&\frac{\{1,0,0,-1\}}{\sqrt{2}},\qquad \bar n=\frac{\{1,0,0,1\}}{\sqrt{2}},
\\\label{L-parametrization:app}
q&=&\{\sqrt{Q^2+\vec q_T^2}\cosh y,|\vec q_T| ,0,\sqrt{Q^2+\vec q_T^2}\sinh y\},
\\\nn
l&=&\{\sqrt{L^2+r_T^2} \cosh \eta,r_T \cos \varphi,r_T \sin \varphi,\sqrt{L^2+r_T^2}\sinh \eta\}.
\end{eqnarray}
The value of $l'$ is fixed by the $\delta$-function in such a way that $l'=q-l$. In these variables the integration measure with the mass-shell $\delta$-functions resolves as
\begin{eqnarray}
&&\int d^4l \delta(l^2)\delta(l'^2)
\\\nn
&&=\int_{0}^\infty dL \int_{0}^\infty dr_T \int_{-\infty}^{\infty} d\eta \int_0^{2\pi} d\varphi \frac{r^2_T}{2Q^2}\delta(L)\delta\(r_T-\frac{Q^2}{2\sqrt{Q^2+\vec q_T^2}\cosh(y-\eta)-2|\vec q_T|\cos\varphi}\).
\end{eqnarray}
Consequently, the integrations over $L$ and $r_T$ are removed using the $\delta$-functions, leaving only the integral over $\eta$ and $\varphi$. The region of integration for $\eta$ and $\varphi$ is restricted by the inequalities (\ref{L:cut-region1:app}), which can be rewritten in terms of the new variables introduced in (\ref{L-parametrization:app}) as
\begin{align}\label{L:cut-region2:app}
&r_T>p_{1T}, && \vec q_T^2+r_T^2-2|\vec q_T|r_T\cos\varphi>p_{2T}^2,
\\\nn
&\eta_{\text{min}}<\eta<\eta_{\text{max}},
&& e^{2\eta_{\text{min}}}<\frac{e^y\sqrt{Q^2+\vec q_T^2}-r_T e^{\eta}}{e^{-y}\sqrt{Q^2+\vec q_T^2}-r_Te^{-\eta}}<e^{2\eta_{\text{max}}}.
\end{align}
The analytical analysis of these boundaries is given in ref.~\cite{Bacchetta:2019sam} (Appendix C).

The final expressions for the cut factors obtained from (\ref{P-general:app}) in the parametrization (\ref{L-parametrization:app}) are
\begin{eqnarray}\label{PU:app}
\mathcal{P}_{U}&=&\int_{\text{cut}} d\eta d\varphi \frac{3}{32 \pi}\frac{2\delta^2+\delta^2\cos2\varphi+(2+\delta^2)\cosh(2 \tilde \eta)-4\delta \sqrt{1+\delta^2}\cos \varphi \cosh \tilde \eta}{[\sqrt{1+\delta^2}\cosh\tilde \eta-\delta \cos\varphi]^4},
\\
\mathcal{P}_0&=&\int_{\text{cut}} d\eta d\varphi \frac{3}{64 \pi}\frac{4+2\delta^2+\delta^2\cos 2\varphi-(2-\delta^2) \cosh(2\tilde \eta)
-4\delta \sqrt{1+\delta^2}\cos\varphi \cosh\tilde \eta}{[\sqrt{1+\delta^2}\cosh\tilde \eta-\delta \cos\varphi]^4},
\\
\mathcal{P}_1&=&\int_{\text{cut}} d\eta d\varphi \frac{-3}{8 \pi}\frac{(\delta \cosh\tilde \eta-\sqrt{1+\delta^2}\cos\varphi)\sinh\tilde \eta
}{[\sqrt{1+\delta^2}\cosh\tilde \eta-\delta \cos\varphi]^4},
\\
\mathcal{P}_2&=&\int_{\text{cut}} d\eta d\varphi \frac{3}{64 \pi}\frac{2\delta^2+(2+\delta^2)\cos2\varphi+\delta^2 \cosh(2\tilde \eta)-4\delta \sqrt{1+\delta^2}\cos \varphi \cosh\tilde \eta
}{[\sqrt{1+\delta^2}\cosh\tilde \eta-\delta \cos\varphi]^4},
\\
\mathcal{P}_3&=&\int_{\text{cut}} d\eta d\varphi \frac{3}{16 \pi}\frac{\sqrt{1+\delta^2}\cos\varphi
-\delta \cosh\tilde \eta}{[\sqrt{1+\delta^2}\cosh\tilde \eta-\delta \cos\varphi]^3},
\\
\mathcal{P}_4&=&\int_{\text{cut}} d\eta d\varphi \frac{3}{16 \pi}\frac{\sinh\tilde \eta
}{[\sqrt{1+\delta^2}\cosh\tilde \eta-\delta \cos\varphi]^3},
\\
\mathcal{P}_5&=&\int_{\text{cut}} d\eta d\varphi \frac{3}{8 \pi}\sin\varphi\frac{\sqrt{1+\delta^2}\cos\varphi-\delta \cosh\tilde \eta
}{[\sqrt{1+\delta^2}\cosh\tilde \eta-\delta \cos\varphi]^4},
\\
\mathcal{P}_6&=&\int_{\text{cut}} d\eta d\varphi \frac{3}{16 \pi}\frac{2\sin\varphi \sinh\tilde \eta}{[\sqrt{1+\delta^2}\cosh\tilde \eta-\delta \cos\varphi]^4},
\\\label{P7:app}
\mathcal{P}_7&=&\int_{\text{cut}} d\eta d\varphi \frac{-3}{16 \pi}\frac{\sin \varphi}{[\sqrt{1+\delta^2}\cosh\tilde \eta-\delta \cos\varphi]^3},
\end{eqnarray}
where $\tilde \eta=\eta-y$, and $\delta=\vec q_T^2/Q^2$. The notation $\int_{\text{cut}}$ implies that the integration over $\eta$ and $\varphi$ is performed in the region restricted by (\ref{L:cut-region2:app}). All these expressions are finite at $\delta\to0$. 

The integrals (\ref{PU:app} -- \ref{P7:app}) are rather cumbersome due to the complexity of both the integration domain and the integrand itself. In order to handle them, one can compute one of the two integrations (either over $\varphi$ or $\eta$) in its indefinite form using the tables of integrals. The second integration is to be evaluated numerically because of its intricacy. In \texttt{artemide}, we first integrate over $\eta$ analytically and then evaluate the remaining integral over $\varphi$ numerically. This order of integration is dictated by the fact that the integrand is smoother along the $\varphi$-direction, while it has exponential behavior along $\eta$. Therefore, the numerical integration procedure over $\varphi$ converges much faster. The boundary values for the integration are solved numerically. An alternative implementation is discussed in ref.~\cite{Bacchetta:2019sam}.

\subsection{Hankel transform}
\label{app:Fourier}

The Hankel transform is a bottleneck for all modern codes specializing in the TMD factorization. The transformation algorithm must be fast (there are $\sim 10^{3}-10^{4}$ calls per single cross-section data point), precise (at least 3-4 digits precision, to match the accuracy of modern experiments and other parts of the theory), and operate in a wide range of $q_T$ ($<1$GeV to $\sim 30-40$GeV). So far, the most productive algorithms were based on the Ogata quadrature \cite{Ogata:2005}, such as those used in refs.~\cite{Moos:2023yfa, Bacchetta:2024qre, Bacchetta:2022awv, Kang:2019ctl}. The inclusion of KPCs imposes a further requirement -- the range of the Hankel transform must reach $q_T\sim Q$ (see appendix \ref{app:convolution}). This is a very demanding requirement because TMD distributions decay as $\sim k_T^{-2}$. The Ogata quadrature has the drawback that its step has to be tuned to the desired scale of $q_T$. Consequently, one cannot (at least in a single run) use Ogata quadrature for a transform with $q_T$ from $0$GeV to $100$GeV (or even larger values, since there are data for $Q\sim 10^3$GeV). Instead, we have implemented a different algorithm based on the Levin system of differential equations \cite{Levin:1982, Levin:1996}. Notably, a very similar algorithm has recently been  developed in ref.~\cite{Diehl:2024mmc} independently. The discussed here algorithm provides a transformation matrix between grids in $b$ and $k_T$-spaces, and is very efficient. It has a fixed precision, which can be increased as needed.

In TMD physics, one deals with the Hankel transform in the following form
\begin{eqnarray}\label{toFourier}
F^{(n)}(k_T)=\frac{M^{2n}}{n!}\int_0^\infty \frac{bdb}{2\pi}\(\frac{b}{k_T}\)^n J_n(b k_T) F(b),
\end{eqnarray}
where $F$ is a TMD distribution, $J_n$ is a Bessel function of the first kind, and $M$ is a constant with dimensions of mass. The order of transformation $n$ is determined by the TMD distribution. In particular, $n=0$ corresponds to the unpolarized distribution, while $n=1$ is for the Boer-Mulders function. 

The Levin approach \cite{Levin:1996} allows to compute finite range integrals with an oscillating integrand of the form
\begin{eqnarray}\label{toFourier:I}
I=\int_{B_a}^{B_b}db\, \vec J(b)\cdot \vec f(b),
\end{eqnarray}
where $\vec J$ and $\vec f$ are vectors (lists) of functions and $\cdot$ indicates the usual scalar product between them. Let the set of functions $\vec J$ be closed under differentiation
\begin{eqnarray}
\frac{d}{d b}\vec J(b)=\mathcal{A}\cdot \vec J(b),
\end{eqnarray}
where $\mathcal{A}$ is a matrix of simple coefficients. Then, it is straightforward to show that
\begin{eqnarray}
I=\int_{B_a}^{B_b}db \frac{d}{d b}\(\vec J(b)\cdot \vec g(b)\)=\vec J(B_b)\cdot \vec g(B_b)-\vec J(B_a)\cdot \vec g(B_a),
\end{eqnarray}
where $\vec g$ satisfies the system of differential equations
\begin{eqnarray}\label{app:Levin-ODE}
\frac{\partial}{\partial b}\vec g(b)+\mathcal{A}^T \cdot \vec g(b)=\vec f(b),
\end{eqnarray}
to be solved. 

In the present case, the integrals (\ref{toFourier}) with $n=0$ and $n=1$ are required. Thus, $\vec J(b)=\{J_0(bk_T),J_1(bk_T)\}$, while the $\mathcal{A}$ matrix is
\begin{eqnarray}
\mathcal{A}=\(\begin{array}{cc}
 \Ds 0&-k_T  \\
 \Ds k_T&-b^{-1} 
\end{array}\).
\end{eqnarray}
The vectors $\vec f$ are $\vec f(b)=\{b F(b),0\}$ for $n=0$, and $\vec f(b)=\{0,b^2 F(b)\}$ for $n=1$ (the factors of $M$ and $k_T$ are multiplied after the integration). Note that the optimal TMD distributions are finite at $b=0$, whereas the evolved TMD distributions vanish at $b=0$. Consequently, the function $\vec f$ vanishes at $b\to0$ as $\vec f(b)\sim b^{n+1}$ at least. 

To solve the system of differential equations (\ref{app:Levin-ODE}) in the range $[B_a,B_b]$, we use the collocation method. As the base function we utilize the Chebyshev interpolant. For that we consider the grid with $N$ nodes distributed as 
\begin{eqnarray}\label{Fourier:node}
b_i=B_b^{\frac{t_i+1}{2}}B_a^{\frac{t_i-1}{2}},\qquad t_i=\cos\(i\frac{\pi}{N}\),
\end{eqnarray}
such that $b_0=B_b$ and $b_N=B_a$. Given the function values at these nodes one can find the interpolation to any internal point with the Chebyshev interpolant (a review of the Chebyshev interpolation properties can be found in ref.~\cite{Diehl:2021gvs}). The derivative of the interpolation function can be found with multiplication by the ``derivative matrix'' $\widehat D_{ij}=D_{ij}/b_i/(B_N-B_0)$ on the vector $f_i=f(b_i)$ with
\begin{eqnarray}
D_{00}=-D_{NN}=\frac{2N^2+1}{6},\qquad D_{ii}=\frac{-t_i}{2(1-t_i^2)},\qquad D_{ij}=\frac{\beta_j}{\beta_i}\frac{(-1)^{i+j}}{t_i-t_j},
\end{eqnarray}
where $\beta_0=\beta_N=1/2$ and $\beta_i=1$ otherwise. The Chebishev polynomials are linearly independent functions, and thus, the approximate solution of (\ref{app:Levin-ODE}) can be found by equalizing the coefficients of each polynomial. It gives rise to a system of linear equations. To write it, we compose the values of the functions at the nodes into the single vectors $G$ and $F$: $G_{i=0,...,N}=g_1(b_i)$, $G_{i=N+1,...,2N+2}=g_2(b_{i-N-1})$, and $F_{i=0,...,N}=f_1(b_i)$, $F_{i=N+1,...,2N+2}=f_2(b_{i-N-1})$. The differential system (\ref{app:Levin-ODE}) is equivalent to
\begin{eqnarray}
S(k_T)G=F,
\qquad \text{with}\qquad
S(k_T)=\(\begin{array}{cc}
  \Ds \widehat{D}   & \Ds k_T I \\
  \Ds -k_T I   & \Ds \widehat{D}-\frac{I}{b}
\end{array}\).
\end{eqnarray}
The block of matrix $S$ is $(N+1)\times (N+1)$ with $I$ being the identity matrix. 

The value of the integral (\ref{toFourier:I}) is 
\begin{eqnarray}
I&=&J_0(k_T B_b)G_0+J_1(k_T B_b)G_{N+1}-J_0(k_T B_a)G_N-J_1(k_T B_a)G_{2N+2}
\\\nn &=&
\(J_0(k_T B_b)~~ +J_1(k_T B_b)~~ -J_0(k_T B_a)~~ -J_1(k_T B_a)\)\cdot\(\begin{array}{c}
    S^{-1}_{0}  \\
    S^{-1}_{N+1}  \\
    S^{-1}_{N}  \\
    S^{-1}_{2N+1}  \\
\end{array}\)\cdot F,
\end{eqnarray}
where $S_i^{-1}$ is the $i$'th line of $S^{-1}(k_T)$, and thus the matrix in the middle is a $4\times (2N+2)$ matrix. Therefore, the computation of the integral (\ref{toFourier:I}) is written as a multiplication by a vector. For a given set of nodes in $k_T$-space, one can precompute these vectors, and form a transformation matrix from a grid in $b$-space to a grid in $k_T$-space. This transformation has a fixed precision, which can be systematically improved by increasing the density of nodes. The method converges better in the integration region where the Bessel function oscillates.

\begin{figure}
\begin{center}
\includegraphics[width=0.99\textwidth]{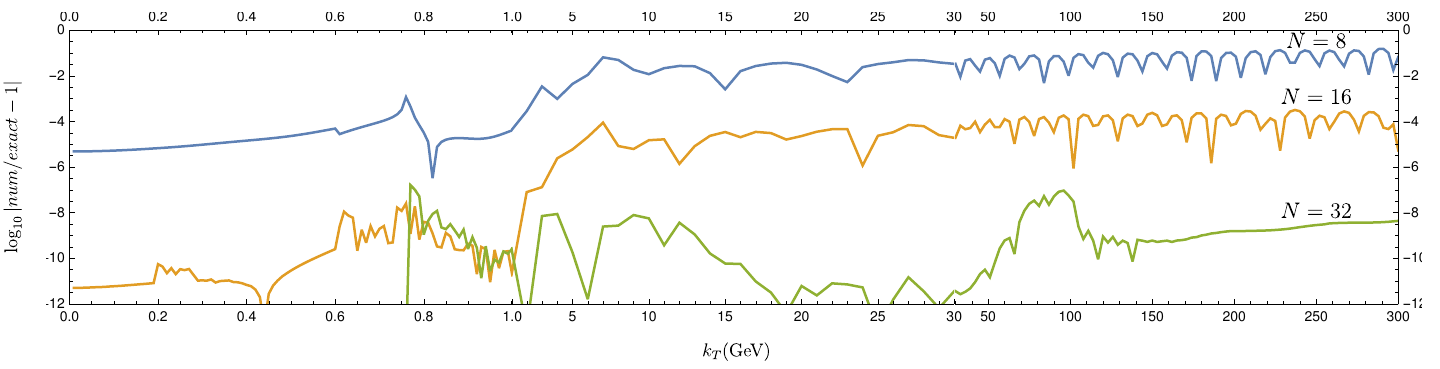}
\caption{\label{fig:Fourier} Comparison of the result of the numeric computation of (\ref{toFourier}) with its exact values. The test function is $f(b)=b^{0.6}\exp(-2.1b)(1+6.3b^2+1.4 b^6)$. The different lines correspond to different number of nodes in sub-grids (indicated in the plot).}
\end{center}
\end{figure}

In order to apply this procedure we split the integral (\ref{toFourier}) into parts
\begin{eqnarray}
F^{(n)}(k_T)=\frac{M^{2n}}{2\pi n! k_T^n}\[\sum_{m=1}^{M} \int_{B_{m-1}}^{B_m}db J_n(b k_T)F(b)+\Delta^{(n)}_0+\Delta^{(n)}_M\],
\end{eqnarray}
where
\begin{eqnarray}\nn
\Delta^{(n)}_0=\int_0^{B_0}db J_n(bk_T)F(b)\simeq \frac{F(B_0)J_{n+1}(B_0k_T)}{k_T},
\qquad
\Delta^{(n)}_M=\int_{B_M}^{\infty}db J_n(bk_T)F(b)\lesssim F(B_M).
\end{eqnarray}
Here, the first estimation is valid for the function vanishing at $b\to0$ at small-$B_0$, and the second estimation is proven in ref.~\cite{Soni:1982}. For each integration interval, we create a sub-grid with nodes (\ref{Fourier:node}), and compute the integral by the Levin method if $B_mk_T<j_{n0}$ ($j_{n0}$ being the position of the first zero of the $J_n$ function); otherwise, we integrate by the Clenshaw-Curtis quadrature (and utilizing the same nodes). The condition $B_mk_T<j_{n0}$ is required to separate the smooth integrals, which are badly convergent with the Levin method. Combining the terms obtained together, we get the result of the integration. 

This method is not adaptive and has a fixed precision. However, by using sufficiently dense grids, any precision can be achieved for a large class of functions. Furthermore, it is a very efficient method, since it provides a transformation matrix between the grids in $b$ and $k_T$ spaces. In our computation we use five sub-grids in $b$-space with $B_m=\{10^{-5},10^{-2},0.2,2.,7.,25.\}$. We also employ a similar Chebyshev-logarithm grid in $k_T$-space with sub-grids in the intervals $\{10^{-2},1.,5.,15.,50.,200.,10^4\}$. Then with $N=16$ we have found that the relative precision of the transformation is $\sim 10^{-4}$ and better. The test has been performed for various kinds of functions. An example of comparison is shown in fig.\ref{fig:Fourier}, for a function with a known exact integral.

\subsection{Convolution in momentum space}
\label{app:convolution}

In the TMD factorization approach the angular structure functions are presented through the convolution integral (\ref{def:convolution}). In this appendix we discuss some details on the structure and implementation of this convolution in  \texttt{artemide}. The integral to consider is
\begin{eqnarray}\label{I0:app}
&&\mathcal{I}[F]=
\\\nn &&4p_1^+p_2^-\int d\xi_1 d\xi_2 \int d^4k_1 d^4k_2 
\delta^{(4)}(q-k_1-k_2)\delta(k_1^2)\delta(k_2^2)
\delta(k_1^+-\xi_1p_1^+)\delta(k_2^--\xi_2p_2^-)
 F,
\end{eqnarray}
where $F$ is a scalar function of $k_{1,2}$, $q$ and $\xi_{1,2}$. All these variables can be expressed in terms of $\vec k_{1,2}^2$ using the $\delta$-functions. In particular, the quarks momentum fractions $\xi_{1,2}$ become
\begin{eqnarray}
\xi_1=\frac{x_1}{2}\(1+\frac{\vec k_1^2}{\tau^2}-\frac{\vec k_2^2}{\tau^2}+\frac{\sqrt{\lambda(\vec k_1^2,\vec k_2^2,\tau^2)}}{\tau^2}\),
\quad
\xi_2=\frac{x_2}{2}\(1-\frac{\vec k_1^2}{\tau^2}+\frac{\vec k_2^2}{\tau^2}+\frac{\sqrt{\lambda(\vec k_1^2,\vec k_2^2,\tau^2)}}{\tau^2}\),
\end{eqnarray}
where $\tau^2=2q^+q^-=Q^2+\vec q_T^2$, $x_1=q^+/p_1^+$, $x_2=q^-/p_2^-$ and $\lambda$ is the triangle function, i.e.,
\begin{eqnarray}
\lambda(a,b,c)=a^2+b^2+c^2-2ab-2ac-2bc.
\end{eqnarray}
The variables $x_{1,2}$ are the LP expressions for the collinear momentum fractions.

\begin{figure}
\begin{center}
\includegraphics[width=0.4\textwidth]{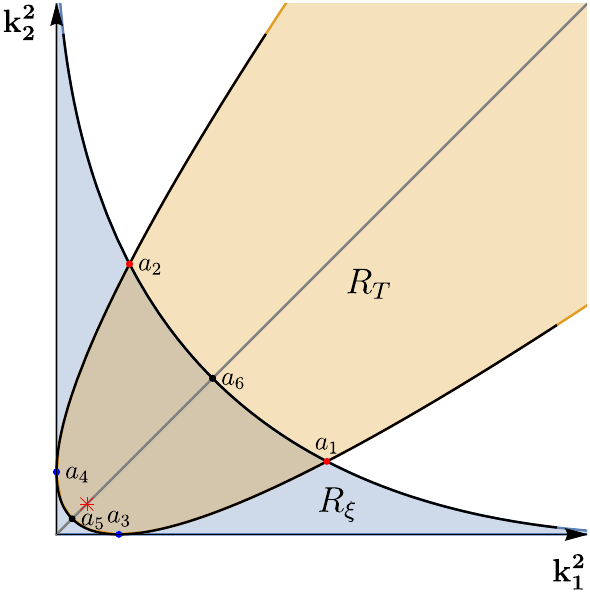}
~
\includegraphics[width=0.45\textwidth]{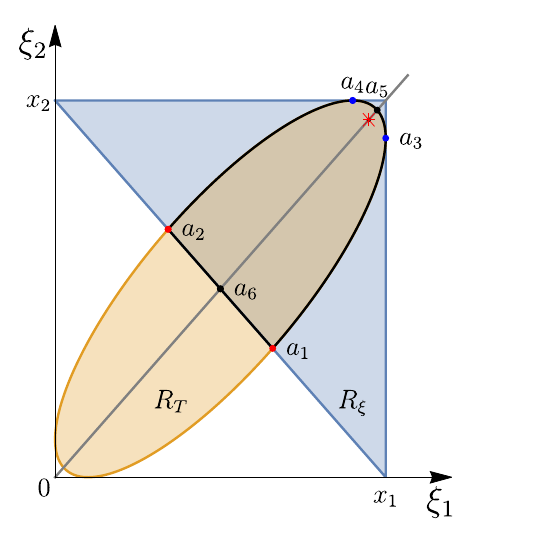}
\caption{\label{fig:PS} The demonstration of the integration regions in the convolution integral (\ref{I0:app}). The left panel shows the regions in $(\vec k_1^2,\vec k_2^2)$ space, while the right panel in $(\xi_1,\xi_2)$ space. The regions $R_\xi$ and $R_T$ are obtained from the independent restrictions as described in the text. The actual integration region is given by the intersection of $R_\xi$ and $R_T$. The values of the utmost points $a_i$ are given in the table \ref{tab:points-a}.}
\end{center}
\end{figure}

The next convenient formulation of the integral (\ref{I0:app}) is 
\begin{eqnarray}\label{I1:app}
&&\mathcal{I}[F]=\int_{R_\xi}   d^2\vec k_1 d^2\vec k_2 
\frac{2}{\sqrt{\lambda(\vec k_1^2,\vec k_2^2,\tau^2)}}\delta^{(2)}(\vec q_T-\vec k_1-\vec k_2)F,
\end{eqnarray}
where $R_\xi$ is the integration region arising due to the restriction $0<\xi_{1,2}<1$. It can be written as
\begin{eqnarray}\label{Rxi:app}
R_\xi:\qquad 0<\vec k_1^2<\tau^2,\qquad 0<\vec k_2^2<(\tau-|\vec k_1|)^2.
\end{eqnarray}
Within this domain, the function $\lambda(\vec k_1^2,\vec k_2^2,\tau^2)>0$. The representation (\ref{I1:app}) is particularly suitable for comparison with the LP computation, since it has the form of the ordinary LP TMD convolution. The LP limit is obtained as
\begin{eqnarray}
&&\lim_{\text{LP}}\mathcal{I}[F]=\frac{2}{Q^2}\int d^2\vec k_1 d^2\vec k_2 
\delta^{(2)}(\vec q_T-\vec k_1-\vec k_2)\lim_{Q\to \infty}F.
\end{eqnarray}
Note that, in the LP limit, the integration boundaries (\ref{Rxi:app}) become infinite, while the values of $\xi_{1,2}$ are set to their LP values, i.e., $\xi_{1,2}\to x_{1,2}$.

The transverse momentum delta-function allows us to reduce the integration to two variables, and imposes extra constraints on the integration region. For example, it is instructive to consider the variables $\vec k_{1,2}^2$ because they are natural arguments of TMD distributions. In these variables the integral turns into
\begin{eqnarray}\label{I2:app}
&&\mathcal{I}[F]=\sum_{s=\pm1}\int_{R_\xi \cap R_T}   d\vec k^2_1 d\vec k^2_2 
\frac{F}{\sqrt{\lambda(\vec k_1^2,\vec k_2^2,\tau^2)}\sqrt{-\lambda(\vec k_1^2,\vec k_2^2,\vec q_T^2)}},
\end{eqnarray}
where $s$ is the relative sign of the orthogonal to $\vec q_T$ components of $\vec k_{1,2}$. The integration limits are further restricted to the area $R_T$
\begin{eqnarray}
R_T:\quad  \vec k_1^2>0,\qquad (|\vec k_1|-|\vec q_T|)^2<\vec k_2^2<(|\vec k_1|+|\vec q_T|)^2.
\end{eqnarray}
Inside the region $R_T$, $\lambda(\vec k_1^2,\vec k_2^2,\vec q_T^2)<0$. The total integration region is given by intersection of $R_\xi$ and $R_T$, and is shown in fig.\ref{fig:PS}. The utmost points of this integration region, denoted by $a_i$, are also marked in fig.\ref{fig:PS}, and their coordinates are summarized in table \ref{tab:points-a}. From this table one can see that the integration range over $\vec k_{1,2}^2$ is restricted such that $\vec k_{1,2}^2\lesssim Q^2/4$, and thus numerically within the factorization domain. Simultaneously, the collinear momentum fractions $x_{1,2}$ are pushed toward the lower values in consistence with the momentum conservation.

\begin{table}
\begin{center}
\renewcommand{\arraystretch}{2.}
\begin{tabular}{l|l|l}
point & $(\vec k_1^2,\vec k_2^2)$ & $(\xi_1,\xi_2)$ 
\\\hline
$a_1$ & $\Ds \Big(\frac{1}{4}(\tau +|\vec q_T|)^2, \frac{1}{4}(\tau -|\vec q_T|)^2\Big) $  
& $\Ds \Big(\frac{x_1}{2}\Big(1+\frac{|\vec q_T|}{\tau}\Big),\frac{x_2}{2}\Big(1-\frac{|\vec q_T|}{\tau}\Big)\Big) $
\\\hline
$a_2$ & $\Ds \Big(\frac{1}{4}(\tau -|\vec q_T|)^2, \frac{1}{4}(\tau +|\vec q_T|)^2\Big) $  
& $\Ds \Big(\frac{x_1}{2}\Big(1-\frac{|\vec q_T|}{\tau}\Big),\frac{x_2}{2}\Big(1+\frac{|\vec q_T|}{\tau}\Big)\Big) $
\\\hline
$a_3$ & $\Ds (\vec q_T^2,0)$  & $\Ds \Big(x_1,x_2\Big(1-\frac{\vec q_T^2}{\tau^2}\Big)\Big) $
\\\hline
$a_4$ & $\Ds (0, \vec q_T^2)$  & $\Ds \Big(x_1\Big(1-\frac{\vec q_T^2}{\tau^2}\Big),x_2\Big) $
\\\hline
$a_5$ & $\Ds \Big(\frac{\vec q_T^2}{4}, \frac{\vec q_T^2}{4}\Big)$  & 
$\Ds \Big(\frac{x_1}{2}\Big(1+\sqrt{1-\frac{\vec q_T^2}{\tau^2}}\Big),\frac{x_2}{2}\Big(1+\sqrt{1-\frac{\vec q_T^2}{\tau^2}}\Big)\Big) $
\\\hline
$a_6$ & $\Ds \Big(\frac{\tau^2}{4}, \frac{\tau^2}{4}\Big)$  & $\Ds \Big(\frac{x_1}{2},\frac{x_2}{2}\Big) $
\\\hline
\end{tabular}
\end{center}
\caption{\label{tab:points-a} Summary on the utmost points of the integration regions shown in fig.\ref{fig:PS}.}
\end{table}

In practice, the integration over $\vec k_{1,2}^2$ is not efficient because the integral can span a rather large area (especially for high-energy experiments), and the integrand has a square-root singularity at the boundary. We have found that it is more convenient to pass to the variables $(\theta,\alpha)$ defined as
\begin{align}
&\vec k_1^2=\frac{\tau^2}{4}\(1-\Lambda+2S+S^2\), && \vec k_2^2=\frac{\tau^2}{4}\(1-\Lambda-2S+S^2\),
\\\nn
& \xi_1=\frac{x_1}{2}(1+S+\sqrt{\Lambda}), && \xi_2=\frac{x_2}{2}(1-S+\sqrt{\Lambda}),
\end{align}
where
\begin{eqnarray}
S=\frac{|\vec q_T|}{\tau}\sin \alpha \cos\theta,\qquad \Lambda=\(1-\frac{\vec q_T^2}{\tau^2}\)\cos^2\alpha.
\end{eqnarray}
The integration region is mapped to a rectangle $\theta\in[0,2\pi]$, and $\alpha\in[0,\frac{\pi}{2}]$. Therefore, the integration takes on a more practical form
\begin{eqnarray}\label{I3:app}
&&\mathcal{I}[F]=\frac{1}{4}\int_0^{\pi/2}d\alpha \int_0^{2\pi}d\theta 
\,F.
\end{eqnarray}
It should be noted that the function $F$ depends non-trivially on $\cos\theta$ due to the non-perturbative TMD distributions. The dependence on $\sin\theta$ is simpler, since it appears only in the kinematic prefactors, in particular, via $(\vec k_1\times \vec k_2)\sim \sin\theta$. Such terms vanish upon integration. The remaining integral over $\theta$ can thus be limited to $2\int_0^\pi$.

Finally, we present the structure functions $\Sigma_n$ (\ref{def:SU}-\ref{def:S7}) in terms of the variables $S$ and $\Lambda$
\begin{eqnarray}
\Sigma_U&=&\frac{4\pi \alpha_{\text{em}}^2}{3N_cs}\sum_{q,G,G'} Q^4\Delta_G^*\Delta_{G'}\,z_{+\ell}^{GG'}z_{+q}^{GG'}\mathcal{C}[1,f_1f_1],
\\
\Sigma_0&=&\frac{4\pi \alpha_{\text{em}}^2}{3N_cs}\sum_{q,G,G'} Q^4\Delta_G^*\Delta_{G'}\Big\{
z_{+\ell}^{GG'}z_{+q}^{GG'}\mathcal{C}[1-\frac{\tau^2\Lambda}{Q^2},f_1f_1]
\\\nn &&
+z_{+\ell}^{GG'}r_{+q}^{GG'}\mathcal{C}[\frac{\tau^2}{4M^2}\(1-S^2-\Lambda-\frac{\tau^2\Lambda}{Q^2}(1+S^2-\Lambda)\),h^\perp_1h^\perp_1]\Big\},
\\
\Sigma_1&=&\frac{4\pi \alpha_{\text{em}}^2}{3N_cs}\sum_{q,G,G'} Q^4\Delta_G^*\Delta_{G'}\Big\{
z_{+\ell}^{GG'}z_{+q}^{GG'}\mathcal{C}[\frac{\tau^2 S\sqrt{\Lambda}}{|\vec q_T|Q},f_1f_1]
\\\nn &&
+z_{+\ell}^{GG'}r_{+q}^{GG'}\mathcal{C}[
\frac{\tau^2 S\sqrt{\Lambda}}{|\vec q_T|Q} \frac{2Q^2-\tau^2(1-S^2+\Lambda)}{4M^2}
,h^\perp_1h^\perp_1]\Big\}
\\\nn
\Sigma_2&=&\frac{4\pi \alpha_{\text{em}}^2}{3N_cs}\sum_{q,G,G'} Q^4\Delta_G^*\Delta_{G'}\Big\{
z_{+\ell}^{GG'}z_{+q}^{GG'}\mathcal{C}[
\frac{Q^2}{\vec q_T^2}\(1-\frac{\tau^2}{Q^2}(1-2S^2+\Lambda)+\frac{\tau^4}{Q^4}\Lambda\),f_1f_1]
\\ && 
+z_{+\ell}^{GG'}r_{+q}^{GG'}\mathcal{C}[
\frac{\tau^2Q^2}{4|\vec q_T|^2M^2}\Big(-1+S^2+\Lambda
\\\nn &&+\frac{\tau^2}{Q^2}(2S^4+(1-\Lambda)^2-3S^2(1+\Lambda))+\frac{\tau^4}{Q^4}\Lambda(1+S^2-\Lambda)\Big)
,h^\perp_1h^\perp_1]\Big\},
\\
\Sigma_3&=&\frac{4\pi \alpha_{\text{em}}^2}{3N_cs}\sum_{q,G,G'} Q^4\Delta_G^*\Delta_{G'}
z_{-\ell}^{GG'}z_{-q}^{GG'}\mathcal{C}[\frac{2S\tau}{|\vec q_T|},\{f_1f_1\}_A]
,
\\
\Sigma_4&=&\frac{4\pi \alpha_{\text{em}}^2}{3N_cs}\sum_{q,G,G'} Q^4\Delta_G^*\Delta_{G'}
z_{-\ell}^{GG'}z_{-q}^{GG'}\mathcal{C}[\frac{2\tau\sqrt{\Lambda}}{Q},\{f_1f_1\}_A]
,
\\
\Sigma_5&=&\frac{4\pi \alpha_{\text{em}}^2}{3N_cs}\sum_{q,G,G'} Q^4\Delta_G^*\Delta_{G'}
i z_{+\ell}^{GG'}r_{-q}^{GG'}
\\\nn &&\qquad\qquad
\mathcal{C}[\frac{-Q\tau^3\sqrt{\Lambda}}{4M^2|\vec q_T|^2}\(1+S^2-\Lambda-\frac{\tau^2}{Q^2}\(1-S^2-\Lambda\)\),\{h^\perp_1h^\perp_1\}_A],
\\
\Sigma_6&=&\frac{4\pi \alpha_{\text{em}}^2}{3N_cs}\sum_{q,G,G'} Q^4\Delta_G^*\Delta_{G'}
i z_{+\ell}^{GG'}r_{-q}^{GG'}
\mathcal{C}[\frac{-\tau^3S}{Q^2|\vec q_T|}\frac{Q^2(-1+S^2-\Lambda)+2\Lambda\tau^2}{4M^2},\{h^\perp_1h^\perp_1\}_A],
\\
\Sigma_7&=&0.
\end{eqnarray}
In the code of \texttt{artemide} these expressions are rewritten in a way that avoids the cancellation of the large numbers $Q^2$ and $\tau^2$, as well as the singularity at $\vec q_T\to 0$.

\bibliography{bibFILE}

\end{document}